\documentclass[a4paper,11pt]{article}
\pdfoutput=1
\usepackage{jheppub}

\usepackage{amsmath}
\usepackage{mathtools}
\usepackage{multirow}
\usepackage{array}

\usepackage{graphicx}
\usepackage{lscape}
\usepackage{subfig}

\usepackage{bbm}
\usepackage{hyperref}
\usepackage{booktabs}

\usepackage[usenames,dvipsnames,svgnames,table]{xcolor}
\usepackage[normalem]{ulem}

\newcommand{\be}{\begin{equation}}
\newcommand{\ee}{\end{equation}}
\newcommand{\bea}{\begin{eqnarray}}
\newcommand{\eea}{\end{eqnarray}}
\newcommand{\bei}{\begin{itemize}}
\newcommand{\eei}{\end{itemize}}
\newcommand{\bean}{\begin{eqnarray*}}
\newcommand{\eean}{\end{eqnarray*}}
\newcommand{\nn}{\nonumber \\}

\def\eps{\epsilon}

\def\top #1{\mathcal{T}_{#1}}

\newcommand\scalemath[2]{\scalebox{#1}{\mbox{\ensuremath{\displaystyle #2}}}}

\newcommand{\xx}{\vec{x}}
\newcommand{\xxi}{\vec{x}_0}
\newcommand{\xxf}{\vec{x}}
\newcommand{\gammaAB}[2]{\ensuremath\gamma}
\newcommand{\gammas}{\ensuremath\gamma_s}
\newcommand{\gammat}{\ensuremath\gamma_t}
\newcommand{\chen}[2]{\ensuremath{\mathcal{C}_{#1}^{\,[#2]}}}

\newcommand{\logxi}[1]{\ensuremath{\log \eta_{#1}(\xxi)}}
\newcommand{\logxf}[1]{\ensuremath{\log \eta_{#1}(\xxf)}}
\newcommand{\logxt}[2]{\ensuremath{\log \eta_{#1}(\xx(#2))}}

\newcommand{\dA}{\ensuremath d\AA}
\newcommand{\pathord}{\ensuremath \mathcal{P}}

\newcommand{\dAk}[1]{\ensuremath \underbrace{\dA \ldots \dA}_{\text{#1 times}}}
\newcommand{\ie}{i.e.\ }

\newcommand{\Den}{\ensuremath D}
\newcommand{\dd}{\ensuremath \mathrm{d}}
\DeclareMathOperator{\dlog}{\mathit{d}log}

\newcommand{\smallvarnothing}{\scalemath{.5}{\varnothing}}
\newcommand{\minus}{\ensuremath \scalebox{0.5}[1.0]{\( - \)}}
\newcommand{\pminus}{\hphantom{\minus}}
\newcommand{\FF}{\ensuremath \text{F}}
\newcommand{\FFvec}{\ensuremath \mathbf{F}}
\newcommand{\GG}{\ensuremath \text{I}}
\newcommand{\GGvec}{\ensuremath \mathbf{I}}
\newcommand{\HH}{\ensuremath \mathbf{H}}
\newcommand{\KK}{\ensuremath \mathbb{K}}
\newcommand{\MM}{\ensuremath \mathbb{M}}
\renewcommand{\AA}{\ensuremath \mathbb{A}}

\title{Two-Loop Master Integrals for the mixed EW-QCD virtual corrections to Drell-Yan scattering}

\author[a]{Roberto Bonciani,}
\author[b]{Stefano Di Vita,}
\author[c,d]{Pierpaolo Mastrolia,}
\author[c]{and Ulrich Schubert}

\emailAdd{roberto.bonciani@roma1.infn.it}
\emailAdd{stefano.divita@desy.de}
\emailAdd{pierpaolo.mastrolia@cern.ch}
\emailAdd{schubert@mpp.mpg.de}

\affiliation[a]{Dipartimento di Fisica, Universit\`a di Roma ``La Sapienza'' and INFN Sezione Roma,\\
 Piazzale Aldo Moro 5, I-00185 Roma, Italy}
\affiliation[b]{DESY, Notkestra\ss e 85, D-22607 Hamburg, Germany}
\affiliation[c]{Max-Planck-Institut f\"ur Physik, F\"ohringer Ring 6, D-80805 M\"unchen, Germany}
\affiliation[d]{Dipartimento di Fisica e Astronomia, Universit\`a di Padova, and INFN Sezione di Padova,\\ Via Marzolo 8, I-35131 Padova, Italy}

\arxivnumber{1604.xxxxx}

\preprint{\texttt{DESY 16-066, MPP-2016-84}}

\keywords{Scattering Amplitudes}

\abstract{
We present the calculation of the master integrals
needed for the two-loop QCD$\times$EW corrections to 
$
q + \bar{q}  \to  l^- + l^+$ and $
q + \bar{q}' \to  l^- + \overline{\nu} \, , 
$
for massless external particles. 
We treat $W$ and $Z$ bosons as degenerate in mass. We identify three types of
diagrams, according to the presence of massive internal lines: the no-mass type, the
one-mass type, and the two-mass type, where all massive propagators, when occurring,
contain the same mass value. We find a basis of 49 master integrals and evaluate
them with the method of the differential equations. The Magnus exponential is
employed to choose a set of master integrals that obeys a canonical system of
differential equations. Boundary conditions are found either by matching the
solutions onto simpler integrals in special kinematic configurations, or by
requiring the regularity of the solution at pseudo-thresholds. The canonical master
integrals are finally given as Taylor series around $d=4$ space-time dimensions, up
to order four, with coefficients given in terms of iterated integrals, respectively
up to weight four. 
}

\begin{document}

\maketitle
\flushbottom

\section{Introduction}

The Drell-Yan production of $Z$ and $W$ bosons \cite{Drell:1970wh} is one of the standard
candles for physical studies at the LHC.
Due to the big cross section and clean experimental signature, Drell-Yan processes  can be
measured  with small experimental uncertainty and, therefore, allow for very  precise tests of
the Standard Model of fundamental interactions (SM). 
They give access to the determination of important parameters of the weak sector, as for
instance the sine of the weak mixing angle and the $W$ boson mass, that together with the top
and the Higgs masses provides stringent constraints on the validity of the SM at the TeV energy
scale.
Furthermore, Drell-Yan processes constitute the SM background in searches of New Physics,
involving  for instance new vector boson resonances, $Z'$ and $W'$, originating from GUT
extensions of the SM.
Finally, the Drell-Yan mechanism is used for constraining parton distribution functions,
for detector calibration and determination of the collider luminosity.
For all these reasons, an accurate and reliable experimental and theoretical control on
Drell-Yan processes would be of the maximum importance for future physics studies at 
colliders.

The theoretical description of Drell-Yan processes currently includes NNLO QCD and
NLO EW radiative corrections, implemented in flexible tools able to provide predictions for
inclusive observables as well as kinematic distributions.
Current theoretical predictions are in good agreement with the experimental measurements.
However, higher theoretical accuracy is needed in order to match the future experimental 
requirements, in particular in view of the run II of the LHC. 
A consistent part of an increasing theoretical accuracy regards higher-order perturbative
corrections.

Very recently, NNNLO QCD corrections were calculated for the Higgs total production cross
section  in gluon-gluon fusion
\cite{Anastasiou:2015ema,Anastasiou:2016cez}. The residual
factorization/renormalization  scales variation moved from about 10-15\% of the  NNLO
calculation (supplemented by NNLL resummation)  to about 5\% of the current result. 
These results will be applied to Drell-Yan as well, since they involve the evaluation of the
same topologies for the calculation of the corresponding Feynman diagrams 
\cite{Gehrmann:2006wg,Heinrich:2007at,Heinrich:2009be,Baikov:2009bg,Lee:2010cga,Gehrmann:2010ue}.

At the same order of accuracy (one can roughly thing to exchange two powers of $\alpha_S$
with one power of $\alpha$), the mixed QCD-EW corrections have to be taken into account.  
As in the case of QCD NNLO with EW NLO perturbative corrections, the mixed QCD-EW corrections 
are expected to become of similar size with respect to QCD NNNLO at high leptonic invariant mass
\cite{Andersen:2014efa}. \\


At LO, the partonic process in the SM is mediated by the exchange of a photon 
or a $Z$/$W$ vector boson, in the $s$ annihilation channel: $q\bar{q} \to \gamma, Z \to l^- l^+$
and $q\bar{q'} \to W \to l \nu$.

At higher orders in the coupling constants, we can distinguish between QCD and electroweak (EW) or mixed (EW-QCD) corrections to the LO process. In the first case, only the initial state 
receives quantum corrections, since the leptonic final state does not couple to gluons.

The NLO QCD corrections to the total cross section were calculated in 
\cite{Altarelli:1979ub,Altarelli:1984pt} and revealed a sizable increase of the cross section 
with respect to the LO result.
The NNLO QCD corrections \cite{Matsuura:1988sm,Hamberg:1990np} stabilized, then, the convergence
of the perturbative series.
%


QCD fixed-order corrections to the total production cross section are supplemented by the
resummation of soft-gluon logarithmically enhanced terms, up to NNNLL approximation 
\cite{Sterman:1986aj,Catani:1989ne,Catani:1990rp,Moch:2005ky}.


EW quantum corrections allow exchanges of quanta between initial and final states. 
Therefore, already at the NLO, massive four-point functions have to be evaluated. 
Since the bulk of the corrections for inclusive observables comes from the resonant region, in
which the exchanged vector boson is nearly on-shell, electroweak NLO corrections to the total
cross section were calculated for the $W$  \cite{Wackeroth:1996hz} and $Z$ \cite{Baur:1997wa}
in narrow-width approximation. 


More exclusive observables are known in the literature. The $Z$ and $W$ production at non-zero
transverse momentum $p_T$ is known at the NLO in QCD 
\cite{Ellis:1981hk,Arnold:1988dp,Gonsalves:1989ar,Brandt:1990vn,Giele:1993dj,Dixon:1998py}
and in the full SM \cite{Kuhn:2004em}. The two-loop QCD helicity amplitudes for the
production of a $Z$ or a $W$
with a photon have also been  calculated \cite{Gehrmann:2011ab}. For small $p_T$ ($p_T \ll
m_W,m_Z$) the convergence of the fixed-order calculation is spoiled by the large logarithmic
terms $\alpha_S^n \log^m{(m_W^2/p_T^2)}$ that have to be resummed 
\cite{Arnold:1990yk,Balazs:1995nz,Balazs:1997xd,Ellis:1997sc,Ellis:1997ii,Qiu:2000ga,Qiu:2000hf,Kulesza:2001jc,Kulesza:2002rh,Landry:2002ix,Bozzi:2010xn}.
Finally, the rapidity distribution of a vector boson is known at the NNLO in QCD
\cite{Anastasiou:2003yy}.

The NLO corrections are available in a fully differential description. They are implemented
in flexible NLO Monte Carlo programs, and merged with QCD parton shower in \texttt{MC@NLO} 
\cite{Frixione:2002ik} and \texttt{POWHEG} \cite{Frixione:2007vw}. In \cite{Barze':2013yca}, the NLO
EW and the QED multiple photon corrections are combined with NLO QCD corrections and parton
shower. 
Pure QED generators are also available 
\cite{CarloniCalame:2003ux,CarloniCalame:2004fza,Bardin:2008fn,Placzek:2013moa}.
Although these implementations provide an accurate description of the process and allow for
realistic phenomenological studies at the hadronic level, they are not accurate enough  for the
performances of the run II at the LHC. 
The NNLO results mentioned above, however, are widely inclusive and they cannot
provide realistic descriptions, that necessarily have to include experimental cuts.
Therefore, a fully differential description of the Drell-Yan process at the NNLO  is
needed. With this respect, the state of the art is represented by the two programs
\texttt{FEWZ} \cite{Melnikov:2006di}, that includes also EW NLO corrections \cite{Li:2012wna}, 
and \texttt{DYNNLO} \cite{Catani:2007vq,Catani:2009sm}. In these two programs, the decay products 
of the vector boson, the spin correlations and the  finite-width effects are also taken 
into account. .

A sizable impact on the $pp(\bar{p}) \to W \to l \nu$ distributions, and therefore
on  the determination of the $W$ mass, comes from the QCD initial state radiation
(ISR)  with QED final state radiation (FSR) or from the real-virtual (factorizable)
corrections. However, at the level of precision required ($\Delta m_W \sim 10~$MeV),
the complete set of mixed QCD-EW corrections may be important and have to be considered. 

The NNLO mixed QCD-EW corrections to the production of a leptonic pair, i.e. order 
$\alpha \alpha_S$ corrections to the LO partonic amplitude, consist 
on two-loop $2\to2$ processes, in which the quark-antiquark initial state goes in 
the final leptonic pair ($l^+l^-$ or $l\nu$), one-loop $2\to3$ processes, in which 
the final leptonic pair is produced together with an unresolved photon or gluon, and 
tree-level $2\to4$ processes in which the leptonic pair is produced together with an 
unresolved photon and an unresolved gluon.

The QCD$\times$QED perturbative corrections were considered in \cite{Kilgore:2011pa}.
In \cite{Kotikov:2007vr}, the mixed two-loop corrections to the form factors for 
the production of a $Z$ boson were calculated analytically, expressing the result in
terms of harmonic polylogarithms and related generalizations.
In \cite{Dittmaier:2014qza}, the authors calculated the mixed corrections in pole
approximation near the resonance region. It particular, they worked out contributions coming
from the QCD corrections to the production and soft-photon exchange between production and
decay process, which cause distortions in the shape of the distributions. In
\cite{Dittmaier:2015rxo}, the factorizable mixed corrections were included in the analysis.
\\

In this article, we present the calculation of the master integrals
(MIs) needed for the virtual corrections to the two-loop $2 \to 2$
processes:
\bea
q + \bar{q}  \to  l^- + l^+ \ , \qquad {\rm and} \quad
q + \bar{q}' \to  l^- + \overline{\nu} \, , \nonumber 
\eea
for massless external particles.
The masses of the $W$ and $Z$ bosons are numerically close to each other, in fact $\Delta
m^2 \equiv m_Z^2 - m_W^2 \ll m_Z^2$. Therefore, in the diagrams containing both $Z$ and $W$
propagators at the same time, one can perform a series expansion in $\xi \equiv \Delta
m^2/m_Z^2 \sim 0.25$.  Within this approximation, all topologies appearing in the two-loop
QCD$\times$EW  virtual corrections to Drell-Yan scattering shall contain either no internal
massive line, or one massive propagator with mass $m_W$, or two massive propagators with the
same mass $m_W$ \cite{Bonciani:2011zz}.  Should they be needed for achieving higher accuracy
within the virtual amplitudes, the coefficients of the series in $\xi$ correspond to scalar
integrals with higher powers of the denominators.

Using the code \texttt{Reduze 2}
\cite{Studerus:2009ye,vonManteuffel:2012np}, the dimensionally
regulated integrals involved in the calculation are reduced to a set
of 49 MIs, which are later determined by means of
the differential equations method
\cite{Kotikov:1990kg,Remiddi:1997ny,Gehrmann:1999as}, reviewed in
\cite{Argeri:2007up,Henn:2014qga}. Of those 49 MIs, 8 contain only 
massless internal lines, 24 involve one massive line and 17 involve two
  massive lines. The system of differential equations obeyed by the
MIs is cast in a canonical form \cite{Henn:2013pwa}, following the
algorithm based on the use of the Magnus exponential, introduced in
\cite{Argeri:2014qva,DiVita:2014pza} \footnote{Other related studies
can be found in \cite{Henn:2013nsa,Gehrmann:2014bfa,Lee:2014ioa}
}.  Boundary conditions are
retrieved either from the knowledge of simpler integrals emerging from
the limiting kinematics, or by requiring the regularity of the
solution at pseudo-thresholds.

Finally, the canonical MIs are given as Taylor series in $\epsilon$ $(=(4-d)/2)$, up to 
order $\eps^4$, being $d$ the dimensional regularization parameter. The coefficients of the
series are pure functions, represented as iterated integrals with rational and irrational
kernels, up to weight four. 
The solution could be expressed in terms of Chen's iterated
integrals. Alternatively, we adopt a {\it mixed representation}, where,
when possible, we make explicit the presence of Goncharov
polylogarithms (GPLs) \cite{Goncharov:polylog,Goncharov2007}, also within the nested integration
structure. 
This representation is suitable for the numerical evaluation of our
solution.

While the two-loop four-point integrals with massless internal lines are
well known in the literature
\cite{Gehrmann:1999as,Smirnov:1999gc,Tausk:1999vh,Anastasiou:2000mf},
the four point integrals with one and two massive internal lines considered here are new and represent the main result of
this communication. \\ 
We verified the numerical agreement of the MIs in the unphysical region against the results of
{\tt SecDec} \cite{Carter:2010hi,Borowka:2012yc,Borowka:2015mxa}.
In particular, because of the presence of irreducible irrational weight functions, 
we found it convenient to cast 5 of the 17 MIs with two massive
internal lines as one-dimensional integral formulas 
\cite{Caron-Huot:2014lda}, involving GPLs in the integrands.
The numerical evaluation of our solutions can, therefore, be performed with the help of the 
{\tt GiNaC} library \cite{Vollinga:2004sn} for the evaluation of GPLs. \\

The article is structured as follows. \\ 
Section \ref{notations} contains our notation and conventions. In Section \ref{sec:diffeq},  we discuss the solution of the canonical differential equations in terms of Chen's iterated
integrals. In section \ref{sec:1mMIs}, we explicitly present the system of differential
equations and the solutions for the one- and two-loop MIs that contain one
massive propagator. In Section \ref{sec:2mMIs}, we give the system of differential equations
for the one-  and two-loop MIs containing two massive propagators. Conclusions
are given In Section \ref{conc}. In Appendix A, we discuss the kinematic domain of our analytic 
results. 
In Appendix B, we give the matrices of the system of differential
equations in canonical form.

Our results are collected in ancillary files, that we include to the \texttt{arXiv} submission.


\section{Notations and Conventions \label{notations}}

In this paper we study the two-loop corrections to the following partonic scattering processes: 
\bea
q(p_1) + \bar{q}(p_2) &\to& l^-(p_3) + l^+(p_4) \, , \label{proc1} \\
q(p_1) + \bar{q}'(p_2) &\to& l^-(p_3) + \overline{\nu}(p_4) \, .
\label{proc2}
\eea
The external particles are considered mass-less and they are on their mass-shell, 
$p_1^2=p_2^2=p_3^2=p_4^2=0$. The scattering can
be described in terms of the Mandelstam variables
\be
s = (p_1+p_2)^2 \, , \qquad t = (p_1-p_3)^2 \, , \qquad u = (p_1-p_4)^2 \, ,
\label{mand}
\ee
in such a way that, for momentum conservation, we have $s+t+u=0$. The physical region is
defined by
\be
s >0 \, , \quad t = - \frac{s}{2} \left( 1- \cos(\theta) \right) \, ,
\ee
where $\theta$ is the scattering angle in the partonic center of mass frame, lying in the range 
$0 < \theta < \pi$. Therefore, while $s>0$, $t$ is always negative and $-s<t<0$.

The quantum corrections to the processes (\ref{proc1}) and (\ref{proc2}) can be expanded in power  series of
the coupling constants. At one loop, the QCD corrections consist on the exchange of a virtual gluon
between the initial-state quarks. The final state is not affected, and at most mass-less three-point
functions have to be evaluated. The EW corrections, instead, consist on the exchange of photons, $Z$
and $W$ bosons. Moreover, these quanta can be exchanged between the quarks in the initial state as
well as the leptons in the final state, but they can also be exchanged between a quark in the initial
state and a lepton in the final state. Consequently, in the calculation of the one-loop corrections
one has to evaluate massive box and vertex diagrams. In the process of $q\bar{q} \to l\nu$ one has to
evaluate diagrams in which a $Z$ and a $W$ bosons are exchanged simultaneously. In order to reduce the
number of scales present in the calculation, we expand the $Z$ propagators around $m_W$:
\be
\frac{1}{p^2-m_Z^2} = \frac{1}{p^2-m_W^2-\Delta m^2} \approx \frac{1}{p^2-m_W^2} + 
\frac{m_Z^2}{(p^2-m_W^2)^2} \, \xi + ...
\ee
where 
\be
\xi = \frac{\Delta m^2}{m_Z^2} = \frac{m_Z^2-m_W^2}{m_Z^2} \sim \frac{1}{4}
\ee
is the effective parameter of the expansion. 
The coefficients of the series in $\xi$ are Feynman diagrams with the
same masses, and eventually with increased powers in the expanded
denominator. Such diagrams depend only on $s,t,$ and one mass $m=m_W$.

However, this does not cause any problem in the calculation, since diagrams with higher
powers of the propagators are in any case reduced to the same set of MIs.
For phenomenological purposes the first order in $\xi$ might be sufficient, but in principle
any order in $\xi$ can be calculated without effort, just relying on the reduction procedure.
We apply the same approximation to the two-loop diagrams as well.

\begin{figure}[t] 
\hspace*{15mm}
\includegraphics[width=0.75\columnwidth,angle=0]{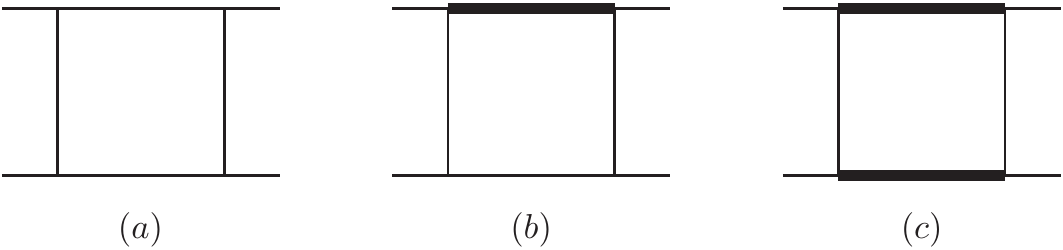} 
\caption{\label{top1} One-loop topologies. Thin lines represent
  massless external particles and propagators, while thick lines
  represent massive propagators.}
\end{figure} 

We calculate the quantum corrections to the processes (\ref{proc1},\ref{proc2}) using a Feynman 
diagrams approach. 
After considering the interference with the leading order, and summing
over the spins and colors, we express the squared absolute value of the amplitude in terms of dimensionally regularized scalar
integrals. 
These integrals are reduced to a set of MIs by means of  integration-by-parts
identities \cite{Tkachov:1981wb,Chetyrkin:1981qh} and Lorentz-invariance identities
\cite{Gehrmann:1999as},  implemented in the computer program\footnote{Other public programs are
available for the reduction to the MIs 
\cite{Anastasiou:2004vj,Smirnov:2008iw,Smirnov:2013dia,Smirnov:2014hma,Lee:2012cn,Lee:2013mka}.} 
\texttt{Reduze 2} \cite{Studerus:2009ye,vonManteuffel:2012np}.

\begin{figure}[t] 
\hspace*{15mm}
\includegraphics[width=0.75\columnwidth,angle=0]{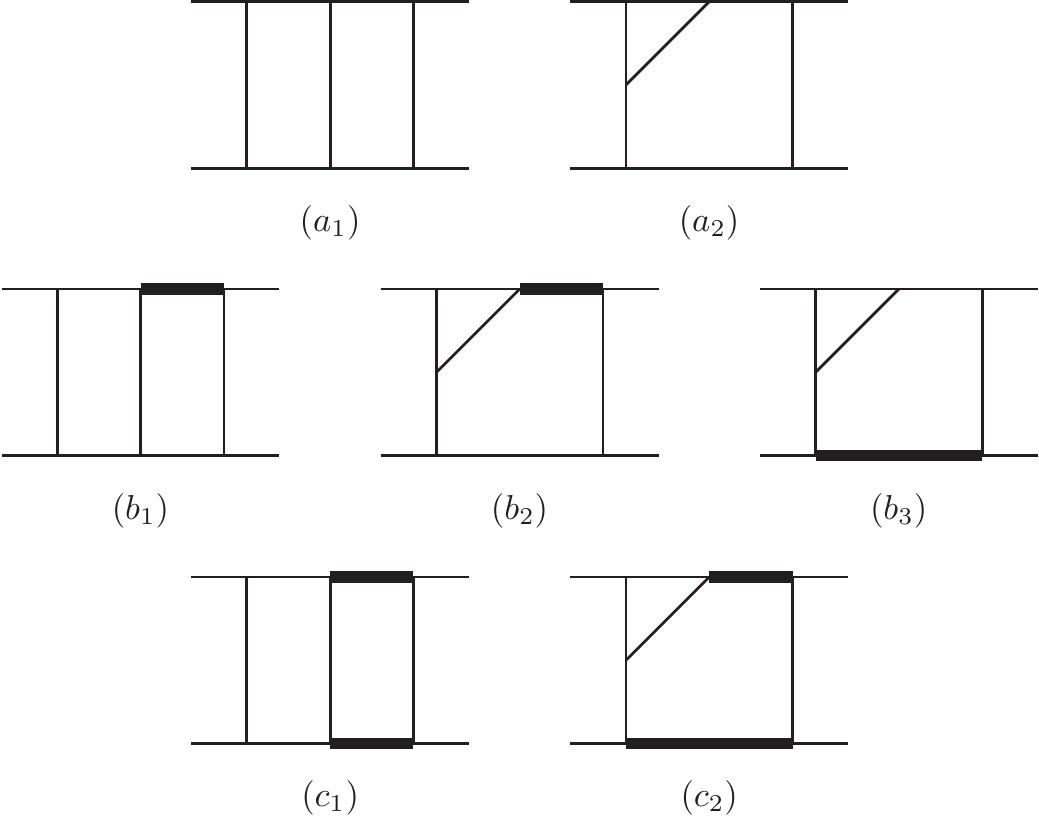} 
\caption{\label{top2} Two-loop topologies. Thin lines represent
  massless external particles and propagators, while thick lines
  represent massive propagators.}
\end{figure}

At one-loop, the topologies involved in the QCD and EW corrections are shown in
figure~\ref{top1}, where we distinguish: 
$a)$ the mass-less case; 
$b)$ the exchange of one massive particle; and $c)$ the exchange of two massive particles. 

At two-loop, the topologies required by the $\mathcal{O}(\alpha
\alpha_S)$ corrections are only planar. They are shown in figure~\ref{top2}.
As for the one-loop case, we consider three classes of diagrams,
according to the presence of massive particles.

Topologies $a_1)$ and $a_2)$ belong to the same 9-denominators mass-less topology.
They reduce to 8 MIs, that were already known in the literature 
\cite{Gehrmann:1999as,Smirnov:1999gc,Tausk:1999vh,Anastasiou:2000mf}. 
Topologies $b_1)$--$b_3)$ have one massive propagator. They reduce to 31 MIs out of which
24 contain one massive propagator and 7 are part of the MIs for topologies $a_1)$ and $a_2)$.
The three-point functions were already known in the literature 
\cite{Fleischer:1998nb,Aglietti:2003yc,Bonciani:2003hc}. 
The four-point functions are calculated and presented here for the first time. 
Topologies $c_1)$ and $c_2)$ have two massive propagators and they reduce to 36 MIs, out of which
17 contain two massive propagators, 15 contain one massive propagator (and they are included in the
set of MIs for topologies $b_1)$--$b_3)$) and 4 contain only massless propagators. 
The three-point functions were known in the literature \cite{Aglietti:2004tq,Aglietti:2004ki} and the four-point functions are presented here for the first time.

The routings for one- and two-mass topologies, at the one- and two-loop level, can be
defined in terms of the following sets of denominators $\Den_n$, where $k_i$ ($i=1,2$) are the 
loop momenta, and $p_i$ ($i=1,\ldots,4$) are the external momenta:

\begin{itemize}

\item\emph{One-mass topologies.}
For the one-loop one-mass integrals (figure~\ref{top1} $b)$, we have:
\begin{gather}
  \Den_1 = k_1^2, \quad
  \Den_2 = (k_1-p_1)^2, \quad
  \Den_3 = (k_1+p_2)^2-m^2, \quad
  \Den_4 = (k_1-p_1+p_3)^2 \nonumber .
\end{gather}
At two loops (figure~\ref{top2} $b_1$--$b_3$), instead, we have:
\begin{gather}
  \Den_1 = k_1^2,\quad
  \Den_2 = k_2^2,\quad
  \Den_3 = (k_1+k_2)^2,\quad
  \Den_4 = (k_1-p_1)^2, \nonumber\\
  \Den_5 = (k_1+p_2)^2,\quad
  \Den_6 = (k_1+k_2-p_1)^2-m^2,\quad
  \Den_7 = (k_1+k_2+p_2)^2,\nonumber\\
  \Den_8 = (k_1+k_2-p_1+p_3)^2,\quad
  \Den_9 = (k_1-p_1+p_3)^2.
\end{gather}

\item\emph{Two-mass topologies.}
For the one-loop two-mass integrals (figure~\ref{top1} $c)$, we have:
\begin{gather}
  \Den_1 = k_1^2, \quad
  \Den_2 = (k_1-p_1)^2-m^2, \quad
  \Den_3 = (k_1+p_2)^2-m^2, \quad
  \Den_4 = (k_1-p_1+p_3)^2 \nonumber .
\end{gather}
At two loops (figure~\ref{top2} $c_1$ and $c_2$), instead, we have:  
\begin{gather}
  \Den_1 = k_1^2,\quad
  \Den_2 = k_2^2,\quad
  \Den_3 = (k_1+k_2)^2,\quad
  \Den_4 = (k_1-p_1)^2, \nonumber\\
  \Den_5 = (k_1+p_2)^2,\quad
  \Den_6 = (k_1+k_2-p_1)^2-m^2,\quad
  \Den_7 = (k_1+k_2+p_2)^2-m^2,\nonumber\\
  \Den_8 = (k_1+k_2-p_1+p_3)^2,\quad
  \Den_9 = (k_1-p_1+p_3)^2.
\end{gather}

\end{itemize}

\noindent
In the following we consider $\ell$-loop Feynman integrals in $\dd$ dimensions, built out of $p$ 
of the above denominators, each raised to some integer power, of the form
\begin{gather}
  \int \widetilde{\dd^d k_1} \ldots \widetilde{\dd^d k_\ell} \, \frac{1}{\Den_{a_1}^{n_1} \ldots \Den_{a_p}^{n_p}},
\end{gather}
where the integration measure is defined as 
\be
\widetilde{\dd^dk_i} \equiv \frac{\dd^d k_i}{(2\pi)^d} \left(\frac{i \, S_\eps}{16 \pi^2} \right)^{-1} \left( \frac{m^2}{\mu^2} \right)^{\eps}, 
\label{eq:intmeasure1}
\ee
with $\mu$ the 't Hooft scale of dimensional regularization, and
\be
S_\eps \equiv (4\pi)^\eps \, \frac{\Gamma(1+\eps)\, \Gamma^2(1-\eps)}{\Gamma(1-2\eps)} \, .
  \label{eq:intmeasure2}
\ee
In eqs.~\eqref{eq:intmeasure1}, \eqref{eq:intmeasure2} we used $\eps=(4-d)/2$.
 	

\section{System of Differential equations for Master Integrals}
\label{sec:diffeq}

In this section, we describe
the general structure of the systems of differential equations obeyed
by the MIs, and the corresponding solutions.  
Sections dedicated to the one-mass and two-mass MIs will follow, where the
details of their complete determination will be provided.

The $b$- and $c$-type MIs are functions of the Mandelstam
invariants defined in eq.~(\ref{mand}) and of the mass $m$.
For their evaluation it is convenient to define the dimensionless
ratios
\begin{gather}
  \label{eq:xyz}
  x \equiv -\frac{s}{m^2}, \quad
  y \equiv -\frac{t}{m^2}, \quad
  z \equiv -\frac{u}{m^2}, \quad \text{ with }\;x+y+z = 0.
\end{gather}
The $b$-type and $c$-type MIs obey systems of partial differential
equations in $x$ and $y$, which can be combined into
matrix equations for their total differentials. In general, the vector
of MIs $\FFvec$ is solution of the following system of differential equation,
\begin{align}
  \label{eq:noncanonicalDE}
d \FFvec = d \KK \, \FFvec \ ,
\end{align}
where the matrix $\KK$ depends \emph{both} on the kinematic variables
and on the spacetime dimension $d=4-2\eps$. 

By means of a suitable basis transformation, built with the help of the
  \emph{Magnus exponential}~\cite{Magnus,Argeri:2014qva} following the
  procedure outlined in Sec.\ 2 of \cite{DiVita:2014pza}, we
  obtain a \emph{canonical} set of MIs~\cite{Henn:2013pwa}. 
  Such a basis obeys a system of differential equation where the dependence on $\eps$ is
  factorized from the kinematics. Moreover, the coefficient matrices can be assembled in a
  (logarithmic) differential form, referred to as canonical
  $\dlog$-form. Hence, the canonical basis $\GGvec$ obeys the following
  system of equations,
\begin{align}
  \label{eq:canonicalDE}
d \GGvec = \eps \, \dA \, \GGvec \ ,
\end{align}
with
\begin{align}
\dA = \sum_{i=1}^n \MM_i \, \dlog \eta_i\,,
\label{dlog-form}
\end{align}
where $\dA $ is the $\dlog$ matrix
written in terms of differentials $\dlog \eta_i$ (that contain the
kinematic dependence) and coefficient matrices $\MM_i$ (with
rational-number entries).  The integrability conditions for
eq.~\eqref{eq:canonicalDE} read
\begin{align}
  \label{eq:canonicalintegrability}
  \partial_a \partial_b \AA - \partial_b \partial_a \AA = 0\,, \qquad [\partial_a \AA,\partial_b \AA]=0\,.
\end{align}

\subsection{General solution}
The general solution of the canonical system of differential
equations~\eqref{eq:canonicalDE} can be compactly written at a point
$\xxf=(x^1,x^2)=(x,y)$ as
\begin{align}
  \label{eq:canonicalsolution}
  \GGvec(\eps,\xxf) = {} & \pathord \exp\left\{\epsilon \int_\gamma \dA\right\} \GGvec(\eps,\xxi)\,,
\end{align}
where $\GGvec(\eps, \xxi)$ is a vector of arbitrary constants, depending
  on $\eps$, while $\dA$ depends only on the kinematic variables.
In the above expression, the \emph{path-ordered} exponential is a
short notation for the series
\begin{align}
  \label{eq:pathorderedexp}
  \pathord \exp\left\{\epsilon \int_\gamma \dA\right\} =
  \mathbbm{1} + \epsilon \int_\gamma \dA + \epsilon^2 \int_\gamma \dA\, \dA + \epsilon^3 \int_\gamma \dA\, \dA\, \dA \ldots\,,
\end{align}
in which the line integral of the product of $k$ matrix-valued 1-forms
$\dA$ is understood in the sense of \emph{Chen's iterated
  integrals}~\cite{Chen:1977oja} (see also~\cite{Brown:2009qja} and
the pedagogical lectures~\cite{Brown:2009lectures}) and $\gamma$ is a
piecewise-smooth path
\begin{align}
     \label{eq:gamma}
     \gammaAB{\xxi}{\xxf}:[0,1]\ni t \mapsto  \gamma(t) = (\gamma^1(t),\gamma^2(t))\,,
\end{align}
such that $\gamma(0)=\xxi$ and $\gamma(1)=\xxf$.  It
follows from Chen's theorem
that the iterated integrals in eq.~\eqref{eq:pathorderedexp} do not
depend on the actual choice of the path, provided the curve does not
contain any singularity of $\dA$ and it does not cross any of its
branch cuts, but only on the endpoints.  In this sense, if one fixes
$\xxi$ and lets $\xxf$ vary, eq.~\eqref{eq:canonicalsolution} can be
thought of as a function of $\xxf$.  In the limit $\xxf\to\xxi$, the
line shrinks to a point and all the path integrals in
eq.\eqref{eq:pathorderedexp} vanish, so that
$\GGvec(\eps,\xxf)\to\GGvec(\eps,\xxi)$, \ie the integration constants have
a natural interpretation as {\it initial} values, and the path-ordered
exponential as {\it evolution} operator. We assume that the vector of
MIs at any point $\GGvec(\xx)$ is normalized in such a way that it admits
a Taylor series in $\epsilon$:
\begin{align}
  \GGvec(\eps,\xx) =  \GGvec^{(0)}(\xx) + \epsilon\, \GGvec^{(1)}(\xx) + \epsilon^2 \GGvec^{(2)}(\xx) + \ldots\,.
\end{align}
The solution $\GGvec(\eps,\xxf)$ is then in principle determined
through~\eqref{eq:canonicalsolution} at any order of the
$\epsilon$-expansion, and reads (up to the coefficient of
$\epsilon^4$)
\begin{align}
  \GGvec^{(0)}(\xxf) = {} & \GGvec^{(0)}(\xxi)\,,
                    \label{eq:epcoeffs0}\\
  \GGvec^{(1)}(\xxf) = {} & \GGvec^{(1)}(\xxi) +  \int_{\gammaAB{\xxi}{P}} \dA \; \GGvec^{(0)}(\xxi)\,,
                    \label{eq:epcoeffs1}\\
  \GGvec^{(2)}(\xxf) = {} & \GGvec^{(2)}(\xxi) +  \int_{\gammaAB{\xxi}{P}} \dA \; \GGvec^{(1)}(\xxi)
                    +  \int_{\gammaAB{\xxi}{P}} \dA\; \dA \; \GGvec^{(0)}(\xxi)\,,
                    \label{eq:epcoeffs2}\\
  \GGvec^{(3)}(\xxf) = {} & \GGvec^{(3)}(\xxi) +  \int_{\gammaAB{\xxi}{P}} \dA \; \GGvec^{(2)}(\xxi)\,,
                    +  \int_{\gammaAB{\xxi}{P}} \dA\; \dA \; \GGvec^{(1)}(\xxi) \nonumber\\
  {} & +  \int_{\gammaAB{\xxi}{P}} \dA\; \dA\; \dA \; \GGvec^{(0)}(\xxi)\,,
       \label{eq:epcoeffs3}\\
  \GGvec^{(4)}(\xxf) = {} & \GGvec^{(4)}(\xxi) +  \int_{\gammaAB{\xxi}{P}} \dA \; \GGvec^{(3)}(\xxi)
                    +  \int_{\gammaAB{\xxi}{P}} \dA\; \dA \; \GGvec^{(2)}(\xxi) \nonumber\\
  {} & +  \int_{\gammaAB{\xxi}{P}} \dA\; \dA\; \dA \; \GGvec^{(1)}(\xxi)
       +  \int_{\gammaAB{\xxi}{P}} \dA\; \dA\; \dA\; \dA \; \GGvec^{(0)}(\xxi)\,.
       \label{eq:epcoeffs4}
\end{align}
The problem of solving~\eqref{eq:canonicalDE}, given a set of initial
conditions $\GGvec(\xxi)$, reduces therefore to the evaluation of
matrices of the type
\begin{align}
  \label{eq:chendAk}
  \int_\gamma \dAk{k}\,,
\end{align}
whose entries, due to~\eqref{dlog-form}, are linear combinations of
Chen's iterated integrals of the form
\begin{align}
  \label{eq:chendlog}
  \int_\gamma \dlog \eta_{i_k} \ldots \dlog \eta_{i_1} \equiv {} & \int_{0\leq t_1 \leq \ldots \leq t_k \leq 1} g^\gamma_{i_k}(t_k) \ldots g^\gamma_{i_1}(t_1) \,dt_1 \ldots \,dt_k\,,
\end{align}
with
\begin{align}
  \label{eq:dlogweight}
  g^\gamma_i(t) = {} & \frac{d}{dt} \log \eta_i(\gamma(t))\,.
\end{align}
We refer to the number of iterated integrations $k$ as the
  \emph{weight} of the path-integral.  The empty integral
(eq.~\eqref{eq:chendlog} for $k=0$) is defined to be equal to 1.
We stress that only the matrices~\eqref{eq:chendAk} do not depend on
the explicit choice of the path. The individual summands of the
form in eq.~\eqref{eq:chendlog}, which contribute to their entries, in
general depend on such a choice.
To keep the notation compact, we
define
\begin{align}
  \chen{i_k,\ldots,i_1}{\gamma} \equiv  \int_\gamma \dlog \eta_{i_k} \ldots \dlog \eta_{i_1} \,,
\end{align}
which also emphasizes that the iterated integrals
in~\eqref{eq:chendlog} are in general \emph{functionals} of the path
$\gamma$.

\subsection{Properties of Chen's iterated integrals}
\label{sec:chen}
The general theory of iterated path integrals was developed by
Chen~\cite{Chen:1977oja}.
Chen's iterated integrals satisfy a number of properties
that we summarize for completeness: 

\begin{itemize}
\item \textit{Invariance under path reparametrization.}  The integral
  $\chen{i_k,\ldots,i_1}{\gamma}$ does not depend on how one chooses
  to parametrize the path $\gamma$.
\item \textit{Reverse path formula.}  If the path $\gamma^{-1}$ is the
  path $\gamma$ traversed in the opposite direction, then
  \begin{align}
    \label{eq:reverse}
    \chen{i_k,\ldots,i_1}{\gamma^{-1}} = (-1)^k \chen{i_k,\ldots,i_1}{\gamma}\,.
  \end{align}
\item\textit{Recursive structure.}  From \eqref{eq:chendlog}
  and~\eqref{eq:dlogweight} it follows that the line integral of one
  $\dlog$ is defined as usual
  \begin{align}
    \label{eq:lineintegral}
    \int_{\gamma} \dlog \eta \equiv {} & \int_{0\leq t \leq 1}  \frac{\dlog \eta(\gamma(t))}{d t} dt\,,
  \end{align}
  and only depends on the endpoints $\xxi,\xxf$
  \begin{align}
    \label{eq:lineintegralres}
    \int_{\gamma} \dlog \eta = \log{\eta(\xxf)} - \log \eta(\xxi)\,.
  \end{align}
  It is convenient to introduce the path integral ``up to some point
  along $\gamma$'': given a path $\gamma$ and a parameter $s\in[0,1]$,
  one can define the 1-parameter family of paths
  \begin{align}
    \label{eq:gammas}
    \gammas:[0,1]\ni t \mapsto \xx = (\gamma^1(s\, t),\gamma^2(s\, t))\,.
  \end{align}
  If $s=1$, then trivially $\gammas=\gamma$.  If $s=0$ the image of
  the interval $[0,1]$ is just $\{\xxi\}$.  If $s\in(0,1)$, then the
  curve $\gammas([0,1])$ starts at $\gamma(0)=\xxi$ and overlaps with
  the curve $\gamma([0,1])$ up to the point $\gamma(s)$, where it
  ends.  It is then easy to see that the path integral along
  $\gamma_s$ can be written as
  \begin{align}
    \label{eq:chens}
    \chen{i_k,\ldots,i_1}{\gammas} = \int_{0\leq t_1 \leq \ldots \leq t_k \leq s} g^\gamma_{i_k}(t_k) \ldots g^\gamma_{i_1}(t_1) \,dt_1 \ldots \,dt_k\,,
  \end{align}
  which differs from eq.~\eqref{eq:chendlog} by the fact that the
  outer integration (\ie the one in $dt_k$) is performed over $[0,s]$
  instead of $[0,1]$.
  Having introduced $\gammas$, we can rewrite~\eqref{eq:chendlog} in a
  recursive manner:
  \begin{align}
    \label{eq:chenrecursive}
    \chen{i_k,\ldots,i_1}{\gamma} = {} & \int_0^1 g^\gamma_{i_k}(s) \, \chen{i_{k-1},\ldots,i_1}{\gammas} ds\,.
  \end{align}
  From eq.~\eqref{eq:chens} we can also immediately derive the
  following identity:
  \begin{align}
    \label{eq:chenderivative}
    \frac{d}{ds}\, \chen{i_k,\ldots,i_1}{\gammas} = g^\gamma_{i_k}(s) \, \chen{i_{k-1},\ldots,i_1}{\gammas}\,.
  \end{align}
  
\item\textit{Shuffle algebra.}  Chen's iterated integrals fulfill
  shuffle algebra relations: if $\vec{m}=m_M,\ldots,m_1$ and
  $\vec{n}=n_N,\ldots,n_1$ (with $M$ and $N$ natural numbers)
  \begin{align}
    \chen{\vec{m}}{\gamma} \, \chen{\vec{n}}{\gamma} = {} &
    \sum_{\text{shuffles }\sigma} \chen{\sigma(m_M),\ldots,\sigma(m_1),\sigma(n_N),\ldots,\sigma(n_1)}{\gamma}\,,
                                                            \label{chen:shuffle}
  \end{align}
  where the sum runs over all the permutations $\sigma$ that preserve
  the relative order of $\vec{m}$ and $\vec{n}$.
    
\item\textit{Path composition formula.}  If
  $\alpha,\beta:[0,1]\to\mathcal{M}$ are such that $\alpha(0)=\xxi$,
  $\alpha(1)=\beta(0)$, and $\beta(1)=\xxf$, then the composed path
  $\gamma\equiv\alpha\beta$ is obtained by first traversing $\alpha$
  and then $\beta$. One can prove that the integral over such a
  composed path satisfies
  \begin{align}
    \label{eq:composed}
    \chen{i_k,\ldots,i_1}{\alpha\beta} = \sum_{p=0}^k \chen{i_k,\ldots,i_{p+1}}{\beta} \, \chen{i_p,\ldots,i_1}{\alpha}\,.
\end{align}

\item\textit{Integration-by-parts formula.}  In order to compute the
  path ordered integral of $k$ $\dlog$ forms using the definition,
  eq.~\eqref{eq:chendlog} (or, equivalently,
  eq.~\eqref{eq:chenrecursive}), in principle one would have to
  perform $k$ nested integrations.  When a fully analytic solution
  cannot be achieved, numerical integration can as well be
  employed. Therefore one can use an alternative form of the Chen
  iterated integral suitable for the combined use of analytic and
  numerical integrations.  In fact, we observe that the innermost
  integration can always be performed analytically using
  \eqref{eq:lineintegral}, so that only $k-1$ integrations are left.
  For instance, in the case $k=2$,
\begin{align}
  \chen{b,a}{\gamma} = {} & \int_0^1 g_{b}(t) \,\chen{a}{\gammat} \,dt \nonumber\\
  = {} & \int_0^1 g_{b}(t) (\logxt{a}{t} - \logxi{a}) \,dt\,.
\end{align}
For $k\geq 3$, one can proceed recursively using eq.~\eqref{eq:chenrecursive},
assuming that the numerical evaluation up to the first $k-1$
iterations is a solved problem.
Using integration by parts, one can show that the numerical
integration over the outermost weight $g_k$ can actually be avoided,
leaving only $k-2$ integrations to be performed
\begin{align}
  \label{eq:chenibpk}
  \chen{i_k,\ldots,i_1}{\gamma} = {} & \logxf{i_k}\, \chen{i_{k-1},\ldots,i_1}{\gamma} - \int_0^1  \logxt{i_k}{t} \, g_{i_{k-1}}(t) \, \chen{i_{k-2},\ldots,i_1}{\gammat} dt\,.
\end{align}

\end{itemize}

\subsection{Mixed Chen-Goncharov representation}
\label{sec:mixedCG}
In principle eq.~\eqref{eq:canonicalsolution} completely determines
the solution, which can be written in terms of Chen's iterated
integrals along an arbitrary piecewise-smooth path (see the discussion below
eq.~\eqref{eq:canonicalsolution}). The initial conditions $\GGvec(\xxi)$
can be computed analytically, if possible, or by means of numerical
methods. The number of iterated integrals that have to be evaluated
numerically can be minimized by the use the of algebraic identities
relating them. According to the
discussion in section~\ref{sec:chen}, the evaluation of the solution
up to weight 4 requires in general 2 nested numerical integrations.

In order to obtain results that allow for an efficient numerical
evaluation, we have chosen to give the solution in a mixed
representation that involves GPLs and general Chen's iterated
integrals.  The representation in terms of GPLs is particularly
convenient because public packages exist, like
\texttt{GiNaC}, that implement their numerical evaluation
in a fast and accurate way.  Whenever the alphabet is rational in the
kinematic variables $x_i$, one can always choose a path that allows to
express the Chen iterated integrals in terms of GPLs, namely the
broken line such that, in each segment, only one of the $x_i$ is
allowed to vary.  Along each segment, by means of factorization over
the complex numbers, one can obtain a linear alphabet and, therefore,
the GPLs representation.  This approach is equivalent to integrating
the differential equations for $x$ and $y$ separately. By integrating,
say, the equation in $x$ one obtains the solution in terms of GPLs of
argument $x$ up to an unknown function $\HH(y)$. By taking the
derivative with respect to $y$ and matching to the equation in $y$,
one obtains a differential equation for $\HH(y)$. The latter can be
again integrated in terms of GPLs of argument $y$, up to a constant.

As we will discuss in section~\ref{sec:2mMIs}, the alphabet for our
differential equations is not always rational in the kinematic
variables we use and, in that case, a representation in terms of GPLs
cannot be given for the complete solution.  To reach the mixed
representation, we have exploited the property of path-independence of
the coefficients of the $\epsilon$-expansion of the solution
eq.~\eqref{eq:canonicalsolution}.  In particular,
eqs.~\eqref{eq:epcoeffs1}-\eqref{eq:epcoeffs4} can be written in an
equivalent alternative form using eq.~\eqref{eq:chenrecursive}:
\begin{align}
  \label{eq:trick}
  \GGvec^{(k)}(\xxf) = {} & \GGvec^{(k)}(\xxi) + \int_0^1 \Bigg[ \frac{\dA(t)}{dt} \GGvec^{(k-1)}(\xx_t) \Bigg] dt\,,
\end{align}
where $\xx_t$ is the point $(x(t),y(t))$ along the curve identified by
$\gamma$.  We see that, in order to build the weight-$k$ coefficient,
one must perform a path integration over the weight-$(k-1)$
coefficient. The choice of such path is independent of the path used
to compute the former because, as we have already discussed, each
coefficient is a function of the sole endpoints.
In other words, as far as the weight-$k$ coefficient of the solution
is concerned, we are free to choose the integration path independently
for each of the $k$ integrations (for each component of $\GGvec(\xx)$).

To see how this can be useful in our computation, we note that the
letters $\eta_i$ (in suitable variables, say $\xx$) can be grouped in
two classes. The first contains the letters that are
rational in the components of $\xx$ and happens to represent the
alphabet for most of the MIs we need to compute.  The second is the
class of letters that are non-rational functions of the variables.
The two classes together constitute the alphabet for the 5 most
complicated MIs.

Starting from the weight-1 coefficient of the solution, we proceed as
follows. As far as the involved $\eta_i$'s belong to the first class
of letters, we can express the solution in terms of
GPLs.
We keep integrating in this manner until, at some weight $k$, the
solution begins to involve non-rational $\eta_i$'s. At this point we
proceed with the path integration as in~\eqref{eq:trick}. Within this
approach, the weight $k-1$ solution is not expressed in terms of
Chen's iterated integrals over an arbitrary path, but in terms of
GPLs. We introduce the following notation to keep our results compact:
{\allowdisplaybreaks[1]
\begin{align}
\chen{a|\vec{m}|\vec{n}}{\gamma} \equiv {} & \int_0^1 \, g_a^\gamma(t) \, G^\gamma_{\vec{m}}(x) \, G^\gamma_{\vec{n}}(y) \,dt \,,\\
\chen{a|\vec{m}|\smallvarnothing}{\gamma} \equiv {} & \int_0^1 \, g_a^\gamma(t) \, G^\gamma_{\vec{m}}(x) \,dt \,,\\
\chen{a|\smallvarnothing|\vec{n}}{\gamma} \equiv {} & \int_0^1 \, g_a^\gamma(t)  \, G^\gamma_{\vec{n}}(y) \,dt \,,\\
\chen{a,\vec{b}|\vec{m}|\vec{n}}{\gamma} \equiv {} & \int_0^1 \, g_a^\gamma(t) \, \chen{\vec{b}|\vec{m}|\vec{n}}{\gammat} \,dt \,,
\end{align}
}
where $G^\gamma_{\vec{m}}(x)$ and $G^\gamma_{\vec{n}}(y)$ stand for the GPLs
$G_{\vec{m}}(x)$ and $G_{\vec{n}}(y)$ evaluated at $(x,y)=(\gamma^1(t),\gamma^2(t))$.

\subsection{Constant GPLs}
In the determination of the boundary values of the MIs we encountered
constant GPLs of argument $1$ with weights drawn from three
sets. For the one-mass MIs there is only one relevant set, with four weights,
\begin{itemize}
\item $\{-1,0,\frac{1}{2},1 \}$\,.
\end{itemize}
For the two-mass MIs we encountered the following two sets, with seven weights each
\begin{itemize}
\item $\{-1,0,-i,i,1,(-1)^{\frac{1}{3}},-(-1)^{\frac{2}{3}} \}$,
\item $\{-1,0,-i,i,1,-(-1)^{\frac{1}{6}},-(-1)^{\frac{5}{6}} \}$,
\end{itemize}
where the former includes the third roots of $-1$ and the latter
involves a subset of the sixth roots of $-1$.  With the help of
\texttt{GiNaC}, we verified that, at order $\eps^k$, the Taylor
coefficient of each MI $\GG_i^{(k)}$ contains a combinations of constant GPLs that 
turns out to be proportional to $\zeta_k$, namely amounting to
$q_{i,k} \ \zeta_k $, with $q_{i,k} \in \mathbb{Q}$.  The
resulting identities were verified at high numerical accuracy.
As examples, we show,
\begin{align}
  0 = {} & G_r + G_{-r^2}\,, \\
  \zeta_2 = {} &   3 G_{0,-r^2}+4 G_{r,-r^2}+4 G_{-r^2,0}-2 G_{-r^2,1}+4 G_{-r^2,r} \nn
    & + 4 G_{-r^2,-r^2}+3 G_{0,r}+4 G_{r,0}-2 G_{r,1}+4 G_{r,r} \,, \\
  - \frac{77 }{8} \zeta_3 = {} &
  G_{-1,-1,\frac{1}{2}}+G_{-1,\frac{1}{2},-1}+G_{-1,\frac{1}{2},1}+3
  G_{0,0,\frac{1}{2}}+3 G_{0,\frac{1}{2},1}+G_{\frac{1}{2},-1,-1} \nn  
  & +
  G_{\frac{1}{2},-1,1}-G_{\frac{1}{2},0,\frac{1}{2}}+4G_{\frac{1}{2},0,1}+G_{\frac{1}{2},1,-1}
  +
  \frac{3}{2} \zeta_2 
  G_{\frac{1}{2}} \, , 
\end{align}
where for simplicity we omitted the argument ($x=1$) of the GPLs and
we defined the weight $r\equiv(-1)^{1/3}$. For related studies see also 
\cite{Broadhurst:1998rz,2007arXiv0707.1459Z,Moriello,Henn:2015sem}.


\section{One-mass Master Integrals}
\label{sec:1mMIs}

In this section we describe the computation of the MIs with one
internal massive line, namely topology $(b)$ of figure~\ref{top1} and
topologies $(b_1)$-$(b_3)$ of figure~\ref{top2}.

\begin{figure}[t]
\begin{center}
\includegraphics[width=0.75\columnwidth,angle=0]{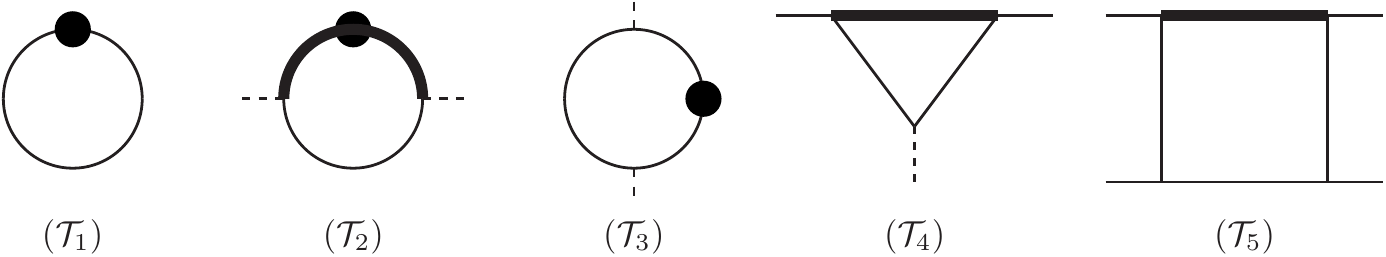} 
\end{center}
\caption{One-loop one-mass MIs $\mathcal{T}_{1,\ldots,5}$. Thin lines
  represent massless external particles and propagators; thick lines
  stand for massive  propagators; an horizontal (vertical) dashed external line
  represents an off-shell leg with squared momentum equal to $s$
  ($t$); dots indicate squared propagators.}
\label{Fig:1M1LMIs}
\end{figure}

\subsection{One-loop}
The following set of MIs for
the one-loop one-mass box obeys a differential equation in $x$ and
$y$, defined in eq.~\eqref{eq:xyz}, which is linear in $\eps$:
\begin{align*}
\FF_1&=\eps \, \top{1}\,,  &
\FF_2&=\eps \, \top{2}\,,  &
\FF_3&=\eps \, \top{3}\,,  \\
\FF_4&=\eps^2 \, \top{4}\,, &
\FF_5&=\eps^2 \, \top{5}\,.  \stepcounter{equation}\tag{\theequation}\label{def:1M1LBasisMIs}
\end{align*}
where the $\mathcal{T}_i$ are depicted in figure~\ref{Fig:1M1LMIs}.
By means of the Magnus exponential~\cite{Magnus,Argeri:2014qva},
according to the procedure outlined in Sec. 2 of
\cite{DiVita:2014pza}, we obtain the canonical MIs
\begin{align*}
\GG_1   &=   \FF_1\,, &
\GG_2   &= -s  \,  \FF_2\,,  &
\GG_3   &= -t \, \FF_3\,,  \\
\GG_4   &= -t \, \FF_4\,, &
\GG_5   &= (s-m^2)  \, t \,  \FF_5 \,. \stepcounter{equation}\tag{\theequation}\label{def:1M1LCanonicalMIs}
\end{align*}
The alphabet of the corresponding $\dlog$-form, eq~\eqref{dlog-form}, is
\begin{align}
\eta_1 & =x\,,&
\eta_2 & =1+x\,, & 
\eta_3 & =y\,, \nn
\eta_4 & =1-y\,, & 
\eta_5 & =x+y\,,&
\label{alphabet:1M1L}
\end{align}
and the coefficient matrices read 
\begin{gather}
  \begin{aligned}
    \MM_1&=
    \scalemath{0.6}{
      \left(
        \begin{array}{ccccc}
          \pminus 0 &\pminus 0 &\pminus 0 &\pminus 0 &\pminus 0 \\
          \pminus 0 &\pminus 1 &\pminus 0 &\pminus 0 &\pminus 0 \\
          \pminus 0 &\pminus 0 &\pminus 0 &\pminus 0 &\pminus 0 \\
          \pminus 0 &\pminus 0 &\pminus 0 &\pminus 0 &\pminus 0 \\
          \pminus 0 &\pminus 2 &\minus  1 &\minus  1 &\pminus 1 \\
        \end{array}
      \right) 
    }\,,&
    \qquad 
    \MM_2&=
    \scalemath{0.6}{
      \left(
        \begin{array}{ccccc}
          \pminus 0 &\pminus 0 &\pminus 0 &\pminus 0 &\pminus 0 \\
          \minus 1  &\minus 2  &\pminus 0 &\pminus 0 &\pminus 0 \\
          \pminus 0 &\pminus 0 &\pminus 0 &\pminus 0 &\pminus 0 \\
          \pminus 0 &\pminus 0 &\pminus 0 &\pminus 0 &\pminus 0 \\
          \pminus 0 &\pminus 0 &\pminus 0 &\pminus 0 &\minus 2 \\
        \end{array}
      \right) 
    }\,,&
    \qquad
    \MM_3&=
    \scalemath{0.6}{
      \left(
        \begin{array}{ccccc}
          \pminus 0 &\pminus 0 &\pminus 0 &\pminus 0 &\pminus 0 \\
          \pminus 0 &\pminus 0 &\pminus 0 &\pminus 0 &\pminus 0 \\
          \pminus 0 &\pminus 0 &\minus  1 &\pminus 0 &\pminus 0 \\
          \pminus 0 &\pminus 0 &\pminus 0 &\pminus 1 &\pminus 0 \\
          \pminus 0 &\pminus 0 &\pminus 0 &\pminus 0 &\minus  1 \\
        \end{array}
      \right)
    }\,,&
  \end{aligned}\nonumber\\
  \begin{aligned}
    \MM_4&=
    \scalemath{0.6}{
      \left(
        \begin{array}{ccccc}
          \pminus 0 &\pminus 0 &\pminus 0 &\pminus 0 &\pminus 0 \\
          \pminus 0 &\pminus 0 &\pminus 0 &\pminus 0 &\pminus 0 \\
          \pminus 0 &\pminus 0 &\pminus 0 &\pminus 0 &\pminus 0 \\
          \minus  1 &\pminus 0 &\pminus 1 &\minus  1 &\pminus 0 \\
          \minus  1 &\pminus 0 &\pminus 1 &\minus  1 &\pminus 0 \\
        \end{array}
      \right) 
    }\,,&
    \qquad 
    \MM_5&=
    \scalemath{0.6}{
      \left(
        \begin{array}{ccccc}
          \pminus 0 &\pminus 0 &\pminus 0 &\pminus 0 &\pminus 0 \\
          \pminus 0 &\pminus 0 &\pminus 0 &\pminus 0 &\pminus 0 \\
          \pminus 0 &\pminus 0 &\pminus 0 &\pminus 0 &\pminus 0 \\
          \pminus 0 &\pminus 0 &\pminus 0 &\pminus 0 &\pminus 0 \\
          \pminus 0 &\minus  2 &\minus  1 &\pminus 1 &\pminus 1 \\
        \end{array}
      \right)
    }\,.
  \end{aligned}
\end{gather}
If $x>0$ and $0<y<1$ all the letters $\eta_i$ are positive.
Since the alphabet is linear in $x$ and $y$, according to the discussion in
section~\ref{sec:mixedCG}, the solution can be conveniently cast
in terms of GPLs.  

Instead of choosing a particular basepoint $\xxi$, the integration
constants of $\GG_{2 \dots 5}$ can be easily fixed by demanding
regularity at the pseudothresholds $ t \rightarrow -m^2 $,
$u \rightarrow 0$, $s \rightarrow 0$ and their reality  in the euclidean region. On
the other hand, $\GG_1$ is a constant and must be determined by direct
integration:
\begin{align}
\GG_{1} &= \frac{\Gamma(1-2\eps)}{\Gamma(1-\eps)^2}\,.
\end{align}

\subsection{Two-loop}
\begin{figure}[t]
\begin{center}
\includegraphics[width=\columnwidth,angle=0]{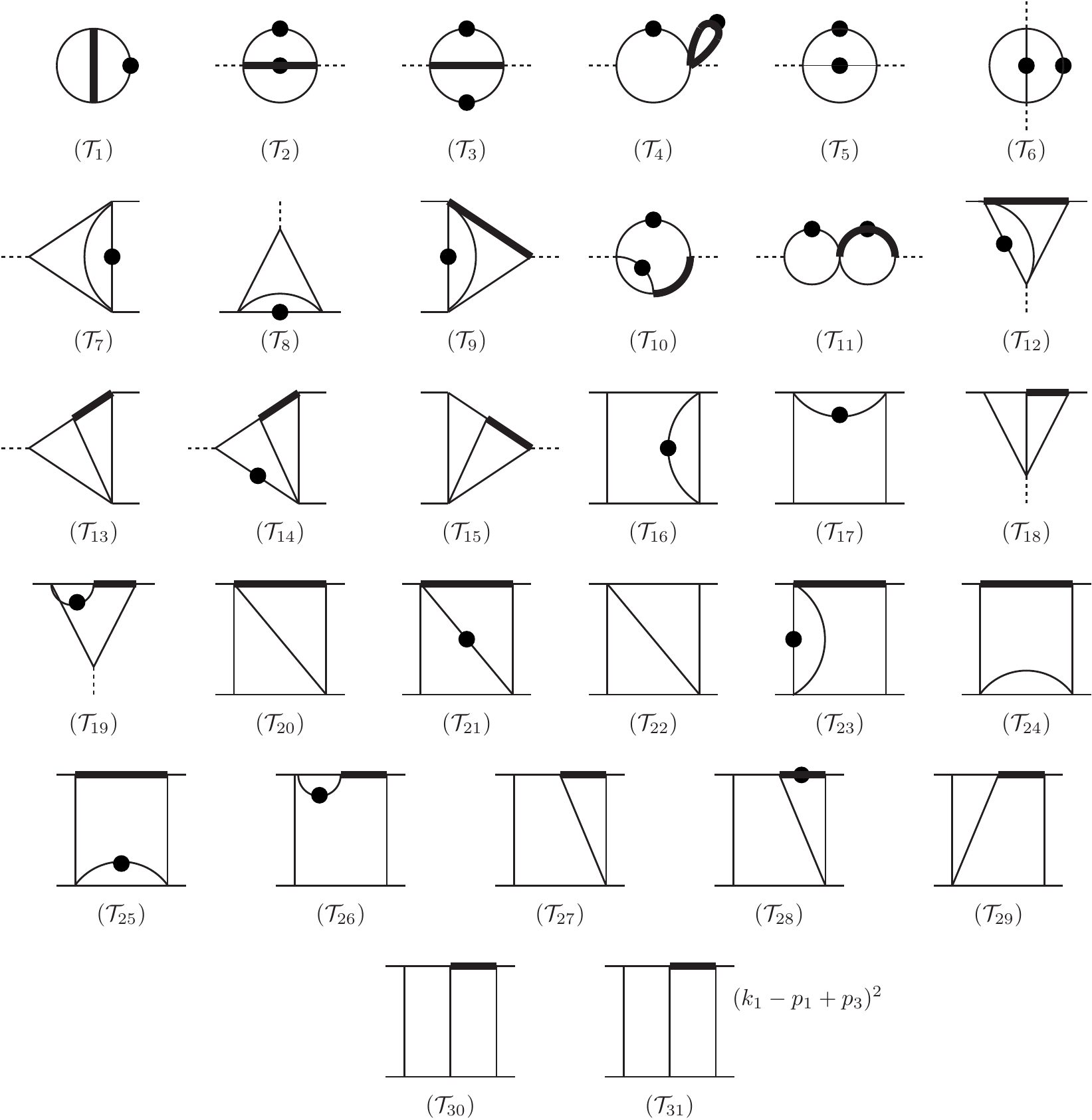} 
\end{center}
\caption{Two-loop one-mass MIs $\mathcal{T}_{1,\ldots,31}$. The
  conventions are as in figure~\ref{Fig:1M1LMIs}.
}
\label{Fig:1M2LMIs}
\end{figure}

At the two-loop order, the following set of MIs admits
$\eps$-linear differential equations in $x$ and $y$ (defined in
eq.~\eqref{eq:xyz}):
\begin{align*}
\FF_1&=(1-\eps)\eps^2 \, \top{1}\,,  &
\FF_2&=\eps^2 \, \top{2}\,,  &
\FF_3&=\eps^2 \, \top{3}\,,  \\
\FF_4&=\eps^2 \, \top{4}\,, &
\FF_5&=\eps^2 \, \top{5}\,,  &
\FF_6&=\eps^2 \, \top{6}\,,  \\
\FF_7&=\eps^3 \, \top{7}\,,   &
\FF_8&=\eps^3 \, \top{8}\,,  &
\FF_9&=\eps^3 \, \top{9}\,,  \\
\FF_{10}&=\eps^2 \, \top{10}\,,  &
\FF_{11}&=\eps^2 \, \top{11}\,,  &
\FF_{12}&=\eps^3 \, \top{12}\,,   \\
\FF_{13}&=\eps^4 \, \top{13}\,,  &
\FF_{14}&=\eps^3 \, \top{14}\,,  &
\FF_{15}&=\eps^4 \, \top{15}\,,  \\
\FF_{16}&=\eps^3 \, \top{16}\,,  &
\FF_{17}&=\eps^3 \, \top{17}\,,  &
\FF_{18}&=\eps^4 \, \top{18}\,,  \\
\FF_{19}&=\eps^3 \, \top{19}\,,  &
\FF_{20}&=\eps^4 \, \top{20}\,,  &
\FF_{21}&=\eps^3 \, \top{21}\,,  \\
\FF_{22}&=\eps^4 \, \top{22}\,,  &
\FF_{23}&=\eps^3 \, \top{23}\,,  &
\FF_{24}&=(1-2\eps)\eps^3 \, \top{24} \,, \\
\FF_{25}&=\eps^3 \, \top{25}\,,  &
\FF_{26}&=\eps^3 \, \top{26}\,,  &
\FF_{27}&=\eps^4 \, \top{27}\,,  \\
\FF_{28}&=\eps^3 \, \top{28}\,,  &
\FF_{29}&=\eps^4 \, \top{29}\,,  &
\FF_{30}&=\eps^4 \, \top{30}\,,  \\
\FF_{31}&=\eps^4 \, \top{31}\,,  \stepcounter{equation}\tag{\theequation}\label{def:1M2LBasisMIs}
\end{align*}
where the $\mathcal{T}_i$ are depicted in figure~\ref{Fig:1M2LMIs}.
Once again,
by means of Magnus exponentials, we are able to obtain a canonical
basis:
\begin{align*}
\GG_1   &=   \FF_1 \,, &
\GG_2   &= -s  \,  \FF_2 \,, &
\GG_3   &= 2 \, m^2 \, \FF_2 + \lambda_- \, \FF_3\,,  \\
\GG_4   &= -s \, \FF_4 \,, &
\GG_5   &= -s  \,  \FF_5  \,,&
\GG_6   &= -t \,     \FF_6 \,, \\
\GG_7   &= -s \, \FF_7 \,, &
\GG_8   &= -t \, \FF_8 \,, &
\GG_9   &= -s \, \FF_9 \,, \\
\GG_{10}&= \frac{m^2 }{2\lambda_+} ( 2 s \lambda_-  \, \FF_{10} -2\,\FF_1 - 3 s \, \FF_{5} ) \,, &
\GG_{11}&= s^2 \, \FF_{11} \,, &
\GG_{12}&= -t  \,    \FF_{12} \,, \\
\GG_{13}&= -s \, \FF_{13} \,, &
\GG_{14}&= s^2 \, \FF_{14} \,, &
\GG_{15}&= -s \, \FF_{15} \,, \\
\GG_{16}&= s \, t\, \FF_{16} \,, &
\GG_{17}&= s \, t \, \FF_{17} \,, &
\GG_{18}&= -t \, \FF_{18} \,, \\
\GG_{19}&= -m^2 \, t\, \FF_{19} \,, &
\GG_{20}&= u \, \FF_{20} \,, &
\GG_{21}&= -t \,\lambda_- \, \FF_{21} \,, \\
\GG_{22}&= u \, \FF_{22} \,, &
\GG_{23}&= -t \, \lambda_- \, \FF_{23} \,, &
\GG_{24}&= -t \, \FF_{24} \,, \\
\GG_{25}&= -t \, \lambda_- \, \FF_{25} \,, &
\GG_{26}&= -t \, m^2 \, (\FF_{17} + \lambda_- \, \FF_{26})\,,  &
\GG_{27}&= s \, t\, \FF_{27} \,, \\
\GG_{28}&= m^2 \, s \, \big( (m^2+t) \, \FF_{28} -2\, \FF_{27} \big) \,, &
\GG_{29}&= (s \, t + m^2 \, u )\, \FF_{29} \,, &
\GG_{30}&= s \, \, t \, \lambda_- \FF_{30} \,, \\
\GG_{31}&= m^2 \, s \, \FF_{29}   - s \, \lambda_- \, \FF_{31} \,, \stepcounter{equation}\tag{\theequation}\label{def:1M2LCanonicalMIs}
\end{align*}
where $\lambda_\pm = \left(m^2 \pm s \right)$. 
   After combining the two differential equations into one
total differential, we find a $\dlog$-form (\ref{dlog-form}) with
the alphabet
\begin{align}
\eta_1 & =x\,,&
\eta_2 & =1+x\,, & 
\eta_3 & =y\,, \nn
\eta_4 & =1-y\,, & 
\eta_5 & =x+y\,,&
\eta_6 & =x+y+xy\,,
\end{align}
which includes the additional letter $\eta_6$ as compared to one-loop
(\ref{alphabet:1M1L}). If $x>0$ and $0<y<1$ all the letters $\eta_i$
are positive. The coefficient matrices are given in the appendix
(\ref{dlog-1M2L}).  Since the additional letter is multilinear in $x$
and $y$, also at the two-loop order we are able to obtain the solution
in terms of GPLs (see the discussion in section~\ref{sec:mixedCG}).

We hereby list the conditions imposed to which integrals for
determining their boundary constants;
\begin{itemize}
\item regularity at $t \rightarrow -m^2 $ and $u \rightarrow 0$ and
  imposing reality on the resulting boundary constants:
  $\GG_{2, \dots, 5,7 \dots 10,12,14 \dots 17,19\dots 31}$\,,
\item limit $s \rightarrow 0$:  $\GG_{11,13}$\,,
\item limit  $t \rightarrow 0$:  $\GG_{18}$\,.
\end{itemize}
 
   This leaves us with $\GG_{1,6}$, to be
determined by direct integration:
\begin{align}
\GG_{1} = {} & -\frac12 \frac{\Gamma (1-2 \eps )^2 \Gamma (1+2 \eps)}{\Gamma (1-\eps)^3 \Gamma (1+\eps)}\,,\\
\GG_{6} = {} & - \frac{y^{-2\eps}}{\pi } \frac{\Gamma \left(\frac{1}{2}-\eps \right) \Gamma \left(\frac{1}{2} + \eps\right) \Gamma (1-2 \eps )}{\Gamma (1-3 \eps ) \Gamma (1+\eps)}\,.
\end{align}

Owing to the explicit representation in terms of GPLs, all the
one-mass MIs can be computed in the whole $(s,t)$ domain (see
appendix~\ref{sec:wz}).  Our results have been successfully checked
against \texttt{SecDec}.

The analytic expressions of all the MIs are explicitly given in
electronic form in ancillary files that can be obtained from the
\texttt{arXiv} version of this paper.


\section{Two-mass Master Integrals}
\label{sec:2mMIs}

In this section we describe the computation of the MIs with two
internal massive lines, namely topology $(c)$ of figure~\ref{top1} and
topologies $(c_1)$-$(c_2)$ of figure~\ref{top2}.

\subsection{One-loop}

\begin{figure}[t]
\begin{center}
\includegraphics[width=0.85\columnwidth,angle=0]{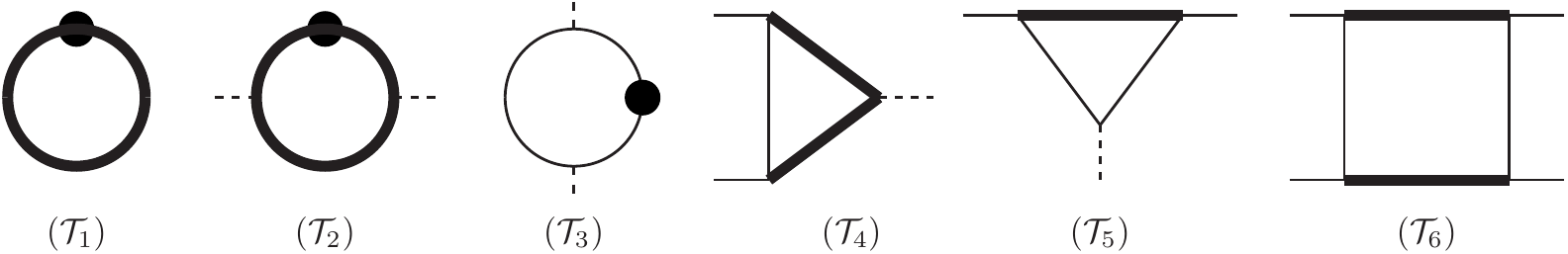} 
\end{center}
\caption{One-loop two-mass MIs $\mathcal{T}_{1,\ldots,6}$. The conventions
  are as in figure~\ref{Fig:1M1LMIs}.}
\label{Fig:2M1LMIs}
\end{figure}

We choose the following set of MIs, admitting a
differential equation linear in~$\eps$
\begin{align*}
\FF_1 &= \epsilon  \, \top{1} \,,  &
\FF_2&=\epsilon \, \top{2} \,, &
\FF_3&=\epsilon  \, \top{3} \,, \\
\FF_4 &= \epsilon^2  \, \top{4} \,,  &
\FF_5&=\epsilon^2 \, \top{5} \,, &
\FF_6&=\epsilon^2 \, \top{6} \,, \stepcounter{equation}\tag{\theequation}\label{def:2M1LBasisMIs}
\end{align*}
where the $\mathcal{T}_i$ are shown in figure~\ref{Fig:2M1LMIs}. After
applying the Magnus transformation we obtain the following canonical basis
\begin{align*}
\GG_1&= \FF_1 \,,&
\GG_2&=- s \, \sqrt{1-\frac{4m^2}{s}}  \, \FF_2 \,,&
\GG_3&= -t \, \FF_3 \,,\\
\GG_4&= -s \, \FF_4 \,,&
\GG_5&= -t \, \FF_5 \,,&
\GG_6&= s \, t \, \sqrt{1 - 4 \frac{m^2}{s}\left(1 +\frac{m^2}{t} \right) } \, \FF_6 \,,\stepcounter{equation}\tag{\theequation}\label{def:2M1LCanonicalMIs}
\end{align*}
The alphabet of the corresponding canonical $\dlog$-form,
\eqref{dlog-form}, is non-rational in $s,t$ and $m$. In particular
four square roots appear
\begin{equation}
\sqrt{-s},\, \sqrt{4m^2-s},\sqrt{-t} ,\, \text{and} \,\,  \sqrt{1 - \frac{4\,m^2}{s}\left(1 +\frac{m^2}{t} \right) }\,.
\label{sqrt:1M2L}
\end{equation}
The latter can be rationalized through the change of variables
\begin{align}
-\frac{s}{m^2}=\frac{(1-w)^2}{w}\,, \qquad -\frac{t}{m^2}=\frac{w}{z} \frac{(1+z)^2}{(1+w)^2}\,.
\label{1loopvartrans}
\end{align}
We note that the above mapping is not invertible at $s=4m^2$. In terms of
  $w$ and $z$, the alphabet reads
\begin{align}
\eta_1 & =z \,, & 
\eta_2 & =1+z \,,&
\eta_3 & =1-z \,, \nn
\eta_4 & =w \,, & 
\eta_5 & =1+w \,,&
\eta_6 & =1-w \,, \nn
\eta_7 & =z-w \,,&
\eta_8 & =z+w^2 \,,&
\eta_{9} & =1-w \, z \,,\nn 
\eta_{10} & =1+w^2 \, z \,,&
\label{alphabet:2M1L}
\end{align}
and the coefficient matrices are 
\begin{gather}
  \begin{alignedat}{4} 
      \MM_1 & =
      \scalemath{0.6}{
      \left(
        \begin{array}{cccccc}
          \pminus 0 & \pminus 0 & \pminus 0 & \pminus 0 & \pminus 0 & \pminus 0 \\
          \pminus 0 & \pminus 0 & \pminus 0 & \pminus 0 & \pminus 0 & \pminus 0 \\
          \pminus 0 & \pminus 0 & \pminus 1 & \pminus 0 & \pminus 0 & \pminus 0 \\
          \pminus 0 & \pminus 0 & \pminus 0 & \pminus 0 & \pminus 0 & \pminus 0 \\
          \pminus 1 & \pminus 0 & \minus  1 & \pminus 0 & \pminus 0 & \pminus 0 \\
          \pminus 0 & \pminus 2 & \pminus 0 & \pminus 0 & \pminus 0 & \pminus 0 \\
        \end{array}
      \right)}\,, &
      \quad 
      \MM_4 & =
      \scalemath{0.6}{
      \left(
        \begin{array}{cccccc}
          \pminus 0 & \pminus 0 & \pminus 0 & \pminus 0 & \pminus 0 & \pminus 0 \\
          \pminus 1 & \pminus 1 & \pminus 0 & \pminus 0 & \pminus 0 & \pminus 0 \\
          \pminus 0 & \pminus 0 & \minus  1 & \pminus 0 & \pminus 0 & \pminus 0 \\
          \pminus 0 & \minus  2 & \pminus 0 & \minus  1 & \pminus 0 & \pminus 0 \\
          \pminus 0 & \pminus 0 & \pminus 0 & \pminus 0 & \pminus 1 & \pminus 0 \\
          \pminus 0 & \pminus 0 & \pminus 0 & \pminus 0 & \pminus 0 & \minus  1 \\
        \end{array}
      \right)}\,, &  
      \quad
      \MM_5 & =
      \scalemath{0.6}{
      \left(
        \begin{array}{cccccc}
          \pminus 0 & \pminus 0 & \pminus 0 & \pminus 0 & \pminus 0 & \pminus 0 \\
          \pminus 0 & \minus  2 & \pminus 0 & \pminus 0 & \pminus 0 & \pminus 0 \\
          \pminus 0 & \pminus 0 & \pminus 2 & \pminus 0 & \pminus 0 & \pminus 0 \\
          \pminus 0 & \pminus 0 & \pminus 0 & \pminus 0 & \pminus 0 & \pminus 0 \\
          \pminus 2 & \pminus 0 & \minus  2 & \pminus 0 & \pminus 0 & \pminus 0 \\
          \pminus 0 & \pminus 0 & \pminus 0 & \pminus 0 & \pminus 0 & \minus  2 \\
        \end{array}
      \right)}\,, &  
      \quad
      \MM_7 & =
      \scalemath{0.6}{
      \left(
        \begin{array}{cccccc}
          \pminus 0 & \pminus 0 & \pminus 0 & \pminus 0 & \pminus 0 & \pminus 0 \\
          \pminus 0 & \pminus 0 & \pminus 0 & \pminus 0 & \pminus 0 & \pminus 0 \\
          \pminus 0 & \pminus 0 & \pminus 0 & \pminus 0 & \pminus 0 & \pminus 0 \\
          \pminus 0 & \pminus 0 & \pminus 0 & \pminus 0 & \pminus 0 & \pminus 0 \\
          \minus  1 & \pminus 0 & \pminus 1 & \pminus 0 & \minus  1 & \pminus 0 \\
          \minus  2 & \pminus 0 & \pminus 2 & \pminus 0 & \minus  2 & \pminus 0 \\
        \end{array}
      \right)}\,,
    \nn
  \end{alignedat} \nn
  \begin{alignedat}{3}
      \MM_8 & =
      \scalemath{0.6}{
      \left(
        \begin{array}{cccccc}
          \pminus 0 & \pminus 0 & \pminus 0 & \pminus 0 & \pminus 0 & \pminus 0 \\
          \pminus 0 & \pminus 0 & \pminus 0 & \pminus 0 & \pminus 0 & \pminus 0 \\
          \pminus 0 & \pminus 0 & \pminus 0 & \pminus 0 & \pminus 0 & \pminus 0 \\
          \pminus 0 & \pminus 0 & \pminus 0 & \pminus 0 & \pminus 0 & \pminus 0 \\
          \pminus 0 & \pminus 0 & \pminus 0 & \pminus 0 & \pminus 0 & \pminus 0 \\
          \pminus 0 & \pminus 0 & \pminus 0 & \pminus 2 & \pminus 2 & \pminus 1 \\
        \end{array}
      \right)}\,, &
      \qquad 
      \MM_9 & =
      \scalemath{0.6}{
      \left(
        \begin{array}{cccccc}
          \pminus 0 & \pminus 0 & \pminus 0 & \pminus 0 & \pminus 0 & \pminus 0 \\
          \pminus 0 & \pminus 0 & \pminus 0 & \pminus 0 & \pminus 0 & \pminus 0 \\
          \pminus 0 & \pminus 0 & \pminus 0 & \pminus 0 & \pminus 0 & \pminus 0 \\
          \pminus 0 & \pminus 0 & \pminus 0 & \pminus 0 & \pminus 0 & \pminus 0 \\
          \minus  1 & \pminus 0 & \pminus 1 & \pminus 0 & \minus  1 & \pminus 0 \\
          \pminus 2 & \pminus 0 & \minus  2 & \pminus 0 & \pminus 2 & \pminus 0 \\
        \end{array}
      \right)}\,, &
      \qquad
      \MM_{10} & =
      \scalemath{0.6}{
      \left(
        \begin{array}{cccccc}
          \pminus 0 & \pminus 0 & \pminus 0 & \pminus 0 & \pminus 0 & \pminus 0 \\
          \pminus 0 & \pminus 0 & \pminus 0 & \pminus 0 & \pminus 0 & \pminus 0 \\
          \pminus 0 & \pminus 0 & \pminus 0 & \pminus 0 & \pminus 0 & \pminus 0 \\
          \pminus 0 & \pminus 0 & \pminus 0 & \pminus 0 & \pminus 0 & \pminus 0 \\
          \pminus 0 & \pminus 0 & \pminus 0 & \pminus 0 & \pminus 0 & \pminus 0 \\
          \pminus 0 & \pminus 0 & \pminus 0 & \minus  2 & \minus  2 & \pminus 1 \\
        \end{array}
      \right)}\,,
  \end{alignedat}
\end{gather}
and $(\MM_2)_{3,3}=-2$ and $(\MM_2)_{5,5}=2$ are the only
non-vanishing entries in $\MM_2$, $(\MM_3)_{6,6}=-2$ is the only
non-vanishing entry in $\MM_3$, and $(\MM_6)_{4,4}=2$ is the only
non-vanishing entry in $\MM_6$. 
In the region $0<w<z<1$ all the letters $\eta_i$ are positive.  For a
detailed discussion of how the interesting regions in the $s,t$ plane
are mapped to the $\mathbb{C}\times\mathbb{C}$ space of the $w,z$
variables, see appendix~\ref{sec:wz}. The alphabet in
(\ref{alphabet:2M1L}) is linear in $z$ but contains letters quadratic
in $w$. As the latter can be linearized by factorization over the
complex numbers, we are once again able to express the solution in
terms of GPLs (see the discussion in
section~\ref{sec:mixedCG}).

The integration constants of $\GG_{4,5,6}$ can be fixed by
requiring their regularity at the pseudothresholds $s \rightarrow 0$,
$t \rightarrow -m^2$ and $u \rightarrow 0$. 
The boundary constant of $\GG_{2}$ can be fixed by taking the
$s \rightarrow 0$ limit. This leaves us with two integrals,
$\GG_{1,3}$, to be determined by direct integration:
\begin{align}
\GG_{1} = {} & \frac{\Gamma(1-2\eps)}{\Gamma(1-\eps)^2} \,,\\
\GG_{3} = {} & \left[ \frac{z}{w} \frac{(1+w)^2} {(1+z)^2} \right]^{\eps} \,.
\end{align}

\subsection{Two-loop}

\begin{figure}[t]
\begin{center}
\includegraphics[width=\columnwidth,angle=0]{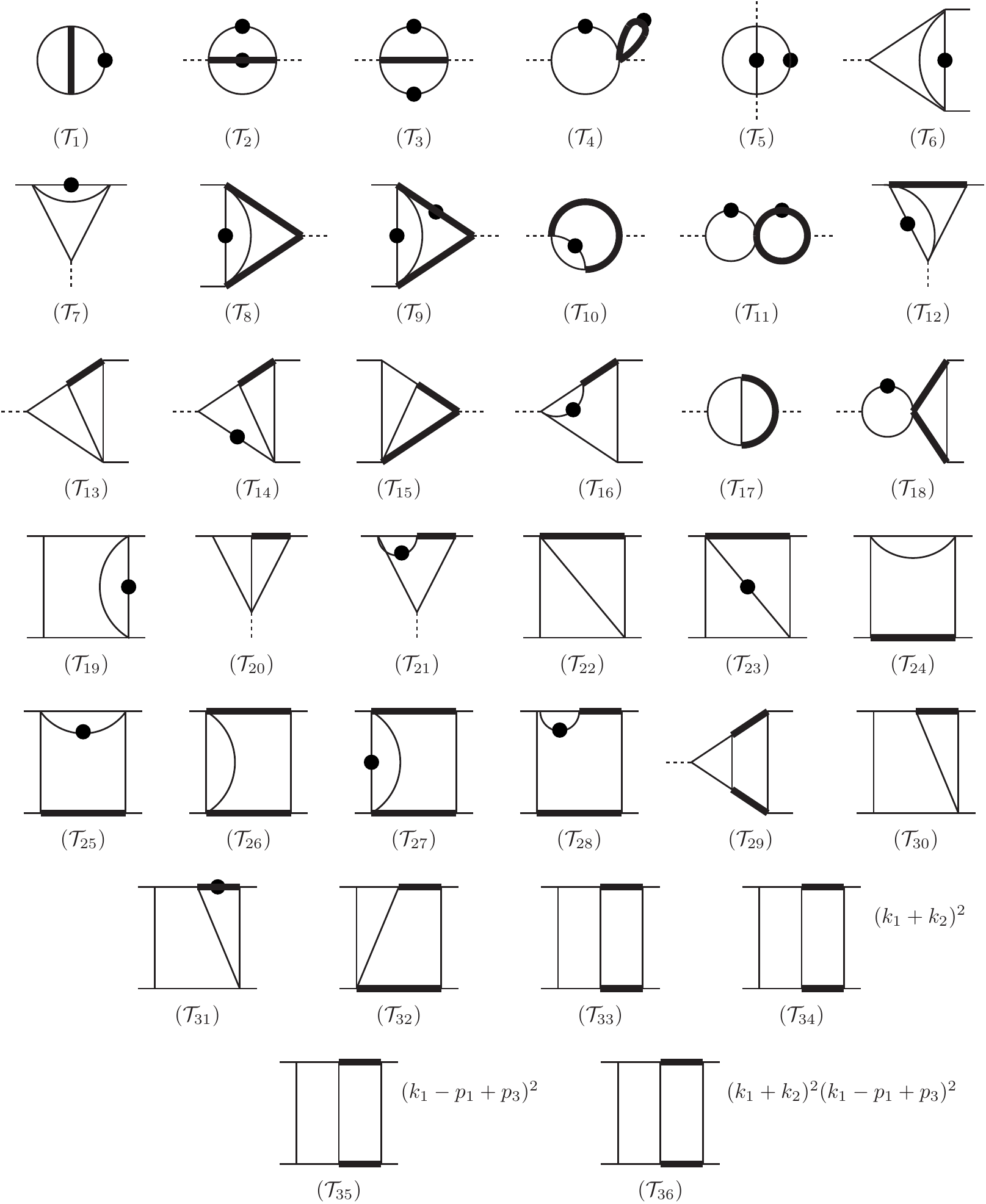} 
\end{center}
\caption{Two-loop two-mass MIs $\mathcal{T}_{1,\ldots,36}$.
  The conventions are as in figure~\ref{Fig:1M1LMIs}.}
\label{Fig:2M2LMIs}
\end{figure}

At the two-loop order we start with the set of MIs 
\begin{align*}
\FF_1 &= (1-\eps) \, \epsilon ^2  \, \top{1} \,,  &
\FF_2&=\epsilon ^2 \, \top{2}  \,,&
\FF_3&=\epsilon ^2 \, \top{3}  \,,\\
\FF_4&=\epsilon ^2 \, \top{4}  \,,&
\FF_5&=\epsilon ^2  \, \top{5}  \,,&
\FF_6&=\epsilon^3 \, \top{6}  \,,\\
\FF_7&=\epsilon ^3 \, \top{7}  \,,&
\FF_8&=\epsilon ^3 \, \top{8}  \,,&
\FF_9&=\epsilon ^2 \, \top{9} \,,\\
\FF_{10}&= (1-2\eps) \, \epsilon ^2 \, \top{10}  \,,&
\FF_{11}&=\epsilon ^2 \, \top{11}  \,,&
\FF_{12}&=\epsilon^3 \, \top{12} \,,\\
\FF_{13}&=\epsilon ^4 \, \top{13}  \,,&
\FF_{14}&=\epsilon ^3 \, \top{14}  \,,&
\FF_{15}&=\epsilon ^4 \, \top{15}  \,,\\
\FF_{16}&=\epsilon ^3 \, \top{16}   \,,&
\FF_{17}&=(1-2\eps) \, \epsilon ^3 \, \top{17} \,,&
\FF_{18}&=\epsilon ^3 \, \top{18}  \,,\\
\FF_{19}&=\epsilon ^3   \, \top{19}   \,,&
\FF_{20}&=\epsilon ^4 \, \top{20}   \,,&
\FF_{21}&=\epsilon ^3 \, \top{21}   \,,\\
\FF_{22}&=\epsilon^4 \, \top{22}  \,,&
\FF_{23}&=\epsilon ^3 \, \top{23}    \,,&
\FF_{24}&= (1-2\eps ) \epsilon ^3 \, \top{24}  \,,\\
\FF_{25}&=\epsilon ^3 \, \top{25}  \,, &
\FF_{26}&=(1-2\eps) \, \epsilon ^3 \, \top{26}  \,, &
\FF_{27}&=\epsilon ^3 \, \top{27}  \,,\\
\FF_{28}&=\epsilon ^3  \, \top{28}   \,,  &
\FF_{29}&=\epsilon ^4 \, \top{29}  \,, &
\FF_{30}&=\epsilon ^4 \, \top{30}  \,,\\
\FF_{31}&=\epsilon ^3 \, \top{31}   \,, &
\FF_{32}&=\epsilon^4 \, \top{32}   \,, &
\FF_{33}&=\epsilon ^4 \, \top{33}  \,,\\
\FF_{34}&=\epsilon ^4 \, \top{34}  \,, &
\FF_{35}&=\epsilon ^4 \, \top{35}  \,, &
\FF_{36}&=\epsilon ^4 \, \top{36}  \,, \stepcounter{equation}\tag{\theequation}\label{def:2M2LBasisMIs}
\end{align*}
where the $\mathcal{T}_i$ are shown in figure~\ref{Fig:2M2LMIs}.
The MIs $\FFvec$ admit $\epsilon$-linear differential equations,
except for one of
them. We have
indeed
\begin{equation}
d\FFvec = d\KK \, \FFvec,\qquad \KK= \KK_{0}+\eps\, \KK_{1}+\frac{1}{1-2\eps} \KK_2 \,,
\end{equation}
where $\KK_0,\KK_1$ and $\KK_2$ do not depend on $\eps$, and $\KK_2$
is non-vanishing only in the inhomogeneous part of the differential
equation for $\FF_{36}$.  In a first step we apply the Magnus
algorithm on $\KK_{0}+\eps\, \KK_{1}$ in order to remove $\KK_{0}$,
and in a second step we apply an ad-hoc transformation in order to
remove the remaining non-linear piece.  

The corresponding canonical
basis reads {\allowdisplaybreaks[1]
  \begin{gather}
    \begin{alignedat}{4}
      \GG_1&= \FF_1 \,,& \qquad
      \GG_2&=-s \, \FF_2 \,,& \qquad
      \GG_3&= m^2 \, (2 \, \FF_2 + \FF_3) -s \, \FF_3 \,,& \qquad
      \GG_4&=-s \, \FF_4 \,,\nn
      \GG_5&=-t \, \FF_5 \,,&
      \GG_6&=-s \, \FF_6 \,,&
      \GG_7&=-t \, \FF_7 \,,&
      \GG_8&=-s \, \FF_8 \,, \nonumber
    \end{alignedat} \\
    \begin{alignedat}{1}
      \GG_9 &= - \sqrt{1-\frac{4 \,m^2}{s}} \left(\frac{3}{2}\, \FF_8 + m^2 \, \FF_9 \right) - \frac{3}{2} s  \, \FF_8 \,, \nn
    \end{alignedat} \\
    \begin{alignedat}{1}
      \GG_{10} = {} & \frac{1}{4} \left(1+ \sqrt{\frac{-s}{4m^2-s}} \right) \left( -2 \FF_1 + (\,m^2-s) \, \left(\FF_2 + \FF_3 \right) +m^2 \, \FF_2 \phantom{\sqrt{1-\frac{4 \, m^2}{s}}} \right .\nn
        & \qquad \qquad \qquad \qquad \qquad \left. + s \, \FF_{10} - s \, \sqrt{1-\frac{4 \, m^2}{s}} \left( \FF_2 + \FF_{10} \right)    \right) \,, \nonumber
    \end{alignedat} \\
    \begin{alignedat}{3}
      \GG_{11}&= s^2 \, \sqrt{1-\frac{4 \,m^2}{s}} \, \FF_{11} \,,& \qquad
      \GG_{12}&=-t \, \FF_{12} \,,& \qquad
      \GG_{13}&=-s \, \FF_{13} \,,\nn
      \GG_{14}&=s^2 \, \FF_{14}\,, &
      \GG_{15}&=-s \, \FF_{15} \,, &
      \GG_{16}&=-m^2 \, s \, \FF_{16} \,, \nn
      \GG_{17}&=-s \, \FF_{17} \,, &
      \GG_{18}&=s^2 \, \FF_{18} \,, &
      \GG_{19}&=s \, t \,  \FF_{19} \,,\nn
      \GG_{20}&= -t \, \FF_{20} \,, &
      \GG_{21}&=-m^2 \, t \, \FF_{21} \,,  &
      \GG_{22}&=u \, \FF_{22} \,, \nn
      \GG_{23}&=(s-m^2) \,t \,  \FF_{23} \,, &
      \GG_{24}&= -t \, \FF_{24} \,, &
      \GG_{25}&=(s-m^2)\, t \, \FF_{25} \,, \nn
    \end{alignedat} \\
    \begin{alignedat}{2}
      \GG_{26}&=-s \, \FF_{26} \,, & \qquad
      \GG_{27}&=s \, t \, \sqrt{1 -  \frac{4 \, m^2}{s}\left(1 +\frac{m^2}{t} \right) } \, \FF_{27} \,, \nn
    \end{alignedat} \\
    \begin{aligned}
      \GG_{28} &=  s \, t \, \sqrt{1 -  \frac{4\, m^2}{s}\left(1 +\frac{m^2}{t} \right) } ( \FF_{25} + m^2 \FF_{28}) + t \, (m^2-s) \, \FF_{25}\,, \nonumber
    \end{aligned} \\
    \begin{alignedat}{2}
      \GG_{29}&=s^2  \, \sqrt{1-\frac{4\,m^2}{s}} \, \FF_{29} \,,& \qquad
      \GG_{30}&=s \, t \, \FF_{30} \,,\nn
      \GG_{31}&=-m^2 \, s \,  (2 \FF_{30}+ (m^2+t ) \, \FF_{31}) \,,&
      \GG_{32}&= s \, t \, \sqrt{1+\frac{m^4}{t^2} - \frac{2\, m^2} {s} \left(1- \frac{u}{t} \right)    }  \, \FF_{32} \,, \nn
      \GG_{33}&=-s^2 \, t \, \sqrt{1 -  \frac{4\, m^2}{s}\left(1 +\frac{m^2}{t} \right) } \, \FF_{33} \,,&
      \GG_{34}&=s^2  \, \FF_{34} \,,
    \end{alignedat} \\
    \begin{alignedat}{1}
      \GG_{35} = {} & s\, \sqrt{1-\frac{4\,m^2}{s}} \left(  2 \, t \, \FF_{32} - s \, t \, \FF_{33} + s \, \FF_{35}  \right)
                   - s^2 \, t \, \sqrt{1 +  \frac{4 \,m^2}{s}\left(1 +\frac{m^2}{t} \right) } \, \FF_{33} \,,\nn
    \end{alignedat} \\
    \begin{aligned}
      \GG_{36} = {} & \frac{s} {2(1-2 \eps )} \, \FF_{17} -  s \, t \, \left( 1- \sqrt{1-\frac{4 \, m^2}{s}} \right) \FF_{32} - s \, t \, \FF_{18} -2 \,t \FF_{22}\\
               & - \frac{2 m^2 \,s }{2 - \frac{s}{m^2} \, (1- \sqrt{1-\frac{4m^2}{s}})} \, (\FF_{29} + t \, \FF_{33} -\FF_{35}) -s \, \FF_{36} \,.
    \end{aligned}
    \label{def:2M2LCanonicalMIs}
  \end{gather}
}%
As compared to the one-loop case (\ref{sqrt:1M2L}) we encounter one
additional square root in the canonical $\dlog$-form
\begin{equation}
\sqrt{1+\frac{m^4}{t^2} - \frac{2\, m^2} {s} \left(1- \frac{u}{t} \right) },
\end{equation}
which is not rationalized by the change of variables in eq.~\eqref{1loopvartrans}.
In terms of $w$ and $z$, the alphabet reads
\begin{gather}
  \begin{alignedat}{3}
    \eta_1 & =z, &\qquad 
    \eta_2 & =1+z,&\qquad
    \eta_3 & =1-z, \nn
    \eta_4 & =w, & 
    \eta_5 & =1+w,&
    \eta_6 & =1-w, \nn
    \eta_7 & =1-w+w^2, & 
    \eta_8 & =1-w \, z,&
    \eta_9 & =z-w, \nn
  \end{alignedat} \\
  \begin{alignedat}{2}
    \eta_{10} & =1+w^2 \, z , &\qquad
    \eta_{11} & =z+w^2,& \nonumber 
  \end{alignedat} \\
  \begin{alignedat}{2}
    \eta_{12}&=4 (1+z)^4 w^3 +(1-w)^2  \, \kappa^2_+(w,z)\,, &\qquad
    \eta_{13} &= (1+w)\sqrt{\rho} + (1-w)\, \kappa_-(w,z)\,, \nn 
    \eta_{14} &= (1+w)\sqrt{\rho} - (1-w)\, \kappa_-(w,z)\,, &\qquad
    \eta_{15} &= (1+w)\sqrt{\rho} + (1-w)\, \kappa_+(w,z)\,, \nn 
  \end{alignedat} \\
  \begin{alignedat}{1}
    \eta_{16} &= \frac{c_1+c_2 \, \sqrt{\rho}}{c_3+c_4 \, \sqrt{\rho}} \,, \nn
  \end{alignedat} \\
  \begin{alignedat}{1}
    \eta_{17} &= 2(1-w)^2 w z^2 + \kappa_-^2(-w,z) + \left(z+w \right) \left(1+w  z\right) \sqrt{\rho} \,,
  \end{alignedat}
  \label{def:2M2Lalphabet}
\end{gather}
where
\begin{align}
  \kappa_\pm(a,b) \equiv a\,(1+b)^2 \pm b\,(1+a)^2\,,
\end{align}
the argument of the square root entering $\eta_{13,\ldots,17}$ is
\begin{align}
  \rho = {} & 4 w z^2 (1+w)^2 - \kappa_+(w,z) \, \kappa_+(-w,-z)\,,
\end{align}
and the four coefficients in $\eta_{16}$ are given by
\begin{align}
  c_1= {} & (1+w)^2 \left(1-4w+w^2 \right) z^2 (1+z)^2 \nn
          & +w^2 (1+z)^6+2 w \left(1-w+w^2\right) z(1+z)^4-2 (1+w)^4 z^3\,, \\
  c_2= {} & \left(1-z^2\right) \kappa_+(w,z)\,, \\
  c_3 = {} & 2 w^8 z^4+2 w^7 z^3 \left(z^2+6 z+1\right)-w^6 (z-1)^2 z^2 \left(z^2+4 z+1\right) \nn
          &-2 w^5 z \left(z^6-z^5-8 z^4-8 z^3-8 z^2-z+1\right) \nn
            & +w^4 \left(z^8-2 z^7-2 z^6+6 z^5-10 z^4+6 z^3-2 z^2-2 z+1\right) \nn
          & -2 w^3 z \left(z^6-z^5-8 z^4-8 z^3-8 z^2-z+1\right) \nn
          & -w^2 (z-1)^2 z^2 \left(z^2+4 z+1\right)+2 w z^3 \left(z^2+6 z+1\right)+2 z^4 \,,\\
  c_4 = {} & -w (1-z^2)(z-w) (1-w z) \left( \kappa_-(-w,-z) + \left(1+w\right)^2 z \right)\,.
\label{def:h}
\end{align}
In the region $0<w<z<1$ all the letters $\eta_i$ are positive.

As already stressed, the alphabet is not rational in $w$ and $z$.
This prevents us from expressing the complete solution in terms of
GPLs. In particular, the structure of the coefficient matrices $\MM_i$
is such that the solution for $\GG_{32}^{(3)}$ and for
$\GG_{32,\ldots,36}^{(4)}$, see
eqs.~\eqref{eq:epcoeffs3}, \eqref{eq:epcoeffs4}, involves path
integration over $\dlog$'s with non rational arguments.  Nevertheless,
the MIs $\GG_{1, \dots, 31}$ admit a representation in terms of GPLs
which is convenient for their numerical evaluation.  As for the
remaining MIs, we followed the procedure outlined in
section~\ref{sec:mixedCG}: we express the solution up to weight 2 for
$\GG_{32}$ and up to weight 3 for $\GG_{33,\ldots,36}$ in terms of
GPLs and then obtain an 1-fold integral representation for the higher
weights (for $\GG_{32, \dots ,36}^{(4)}$ we use eq.~\eqref{eq:chenibpk}).

We hereby list the conditions imposed to integrals $\GG_{1,\ldots,31}$
for determining their boundary constants;
\begin{itemize}
\item independent input: $\GG_{1,4, \dots, 7 ,13 ,14 ,20,25}$\,,
\item regularity at $s \rightarrow 0$: $\GG_{24}$\,,
\item regularity at $t \rightarrow -m^2$: $\GG_{12,21}$\,,
\item regularity at $u \rightarrow 0$:  $\GG_{19,30,31}$\,,
\item limit $s \rightarrow 0$: $\GG_{2,3,6, \dots, 10,15 \dots 18,29}$\,,
\item limit $t \rightarrow -m^2$ and $s \rightarrow 0$: $\GG_{28}$\,,
\item regularity at $s \rightarrow 0$ and matching to independent input:  $\GG_{22,23}$\,.
\end{itemize} 
For the MIs $\GG_{32,\ldots,36}$ we observe that regularity at
$u=s=t=0$, corresponding to $\xx_0=(w_0,z_0)=(1,-1)$, implies
\begin{align}
\GG_{32, \dots, 36}(\eps,\xx_0)=0 \,,
\label{boundary:Ints32to36}
\end{align}
that we choose as initial condition of our solution in terms of iterated
integrals.

The MIs $\GG_{1,\ldots,31}$ are represented in terms of GPLs, and
can be computed on the whole $(s,t)$ plane (except for the line
$s=4m^2$, see appendix~\ref{sec:wz} for further comments).

The explicit evaluation of $\GG_{32,\ldots,36}$ requires a careful choice of the 
integration path, in such a way that no branch cuts are crossed.
We successfully checked our results in the unphysical region $s<0$ (see
appendix~\ref{sec:wz}) against the numerical values  
obtained with \texttt{SecDec}. The evaluation of our
analytic result relies on the use of \texttt{GiNaC} for the
computation of the GPLs and on a one-dimensional integration for the
cases where non-rational weights appear in the most external
iteration, according to the eq.~\eqref{eq:chenibpk}. As for the
latter, we exploited the propriety of path-independence to choose
simple paths (that avoid the singularities on the way from the
basepoint to the chosen endpoints).  Let us remark that in this work
we did not focus on the the computing performances of the numerical
evaluation of the mixed Chen-Goncharov iterated integrals appearing in
our analytic expression. This aspect, together with a study of the
analytic properties of our solutions in the whole phase-space,
requires a dedicated future investigation.

The analytic expressions of all the MIs are explicitly given in
electronic form in ancillary files that can be obtained from the
\texttt{arXiv} version of this paper.
 	

\section{Conclusions \label{conc}}
In this article, we presented the calculation of the master integrals (MIs)
needed for the
virtual QCD$\times$EW two-loop corrections to the Drell-Yan scattering
processes,
$$
q + \bar{q}  \to  l^- + l^+ \, \ , \qquad 
q + \bar{q}' \to  l^- + \overline{\nu} \, , 
$$
for massless external particles.  Besides the exchange of massless
gauge bosons, such as gluons and photons, the relevant Feynman
diagrams involve also the presence of $W$ and $Z$ propagators.  Given
the small difference between the masses of the $W$ and $Z$ bosons, in
the diagrams containing both virtual particles at the same time, we
performed a series expansion in the difference of the squared
masses. Owing to this approximation, we distinguished three types of
diagrams, according to the presence of massive internal lines: the
no-mass type, the one-mass type, and the two-mass type, where all
massive propagators, when occurring, contain the same mass value.  The
evaluations of the four point functions with one and two internal
massive propagators are the main novel results of this communication.

To achieve it, we identified a basis of 49 MIs and evaluated them with
the method of the differential equations. With the help of the Magnus
exponential, the MIs were found to obey a canonical system of
differential equations.  Boundary conditions were imposed either by
matching the solutions onto simpler integrals in special kinematic
configurations, or by requiring the regularity of the solution at
pseudo-thresholds.  The canonical MIs were given as Taylor series
around $d=4$ space-time dimensions, up to order four, whose
coefficients were given in terms of iterated integrals up to weight
four.  The solution could be expressed in terms of Chen's iterated
integrals, yet, we adopted a mixed representation in terms of
Chen-Goncharov iterated integrals, suitable for their numerical
evaluation. Further studies concerning the analytic properties of the
presented MIs in the whole phase-space, and the optimization of their
numerical evaluation will be the subject of a forthcoming publication.


\acknowledgments 

We would like to thank Valery Yundin for his contribution during the
early stages of this project.
We thank Lorenzo Tancredi for clarifying discussions and Matthias
Kerner for technical support on \texttt{SecDec}.
Some of the algebraic manipulations required in this work were carried
out with {\tt FORM}~\cite{Kuipers:2012rf}. Some of the Feynman
diagrams were generated by FeynArts~\cite{Kublbeck:1990xc,Hahn:2000kx}
and drawn with {\tt Axodraw}~\cite{Vermaseren:1994je}.
The work of R.B.\ was partly supported by European Community Seventh
Framework Programme FP7/2007-2013, under grant agreement N.302997.
The work of P.M.\ and U.S.\ was supported by the Alexander von Humboldt
Foundation, in the framework of the Sofja Kovalevskaja Award 2010,
endowed by the German Federal Ministry of Education and Research.
R.B.\ and S.D.V.\ would like to thank the Galileo Galilei Institute for
Theoretical Physics for hospitality during the initial part of this
work.

\appendix
\section{Variables for the one-mass and two-mass integrals}
\label{sec:wz}

In this section we discuss the domain of the variables employed in the analytic 
expressions of the MIs for the Drell-Yan process, both in the case with one massive 
propagator and in the one with two massive propagators.

\subsection{One-mass type}

For the evaluation of the one-mass MIs we simply rescale by the squared mass the 
Mandelstam invariants. All the analytic results are given in terms of two-dimensional
generalized polylogarithms, functions of the variables
\be
x = - \frac{s}{m^2} \, , \quad y = - \frac{t}{m^2} \, .
\ee
In the unphysical region $s<0$, $x$ is real and positive. Correspondingly, $y$ can be
either positive or negative.

The analytic continuation to the physical region requires the Feynman prescription on
the invariants. There, $s$ becomes positive, with a positive vanishing imaginary part,
$s+i0^+$. Accordingly, $x$ is
negative: 
\be 
x \to - x' - i0^+ \, , 
\ee 
with 
\be 
x' = \frac{s}{m^2} > 0 \, .  
\ee
On the other hand, $t$ is negative (with a positive vanishing imaginary part) and ranges
between 0 and $-s$, $-s<t<0$.

The numeric evaluation of the MIs expressed in terms of GPLs of the variables $x$ and
$y$ can be done in the whole $s,t$ plain using the routines in \cite{Vollinga:2004sn}
expressing our analytic formulas in terms of GPLs evaluated in 1 and giving the explicit
imaginary part to the Mandelstam variables (see for instance \cite{Bonciani:2010ms}).

\subsection{Two-mass type}

For the evaluation of the two-mass MIs, see section~\ref{sec:2mMIs}, we
find it convenient to introduce the reduced variables $w$ and $z$
defined by 
\be
\label{eq:wz}
- \frac{s}{m^2} = \frac{(1-w)^2}{w} \, , \qquad 
- \frac{t}{m^2} = \frac{w}{z} \frac{(1+z)^2}{(1+w)^2} \, .
\ee

We note that the above mapping allows the evaluation of our results everywhere in the
$(s,t)$ plane, with the exception of the value $w=-1$ (corresponding to $s=4m^2)$. For
that specific value of $w$, the $t$ dependence in $z$ gets lost by construction, and 
$z=-1$ independently on $t$.

The evaluation of the solution at $s=4m^2$ requires further investigations and it
will be addressed in a forthcoming publication. 

\subsubsection{Range of values for $w$}
For $w$, defined by the first of eqs.~\eqref{eq:wz}, we choose the following root:
\be
\label{eq:w}
w = \frac{\sqrt{4m^2-s-i0^+}-\sqrt{-s-i0^+}}{
\sqrt{4m^2-s-i0^+}+\sqrt{-s-i0^+}} \, , 
\ee
where we explicitly used the Feynman prescription $s+i0^+$.

\begin{enumerate}
\item If $s<0$, we have positive $w$ and $0<w<1$. In particular, when
  $s \to - \infty$, $w \to 0$, while for $s \to 0$, $w \to 1$. 
\item If $0<s<4m^2$, $w$ becomes a phase. In fact \be
  w = \frac{\sqrt{4m^2-s}+i\sqrt{s}}{
    \sqrt{4m^2-s}-i\sqrt{s}} = e^{i \phi} \, , 
  \ee
  where
  \be
  \phi = 2 \arctan{\sqrt{\frac{s}{4m^2-s}}}
  \ee
  and $0<\phi<\pi$.
\item If $s>4m^2$, $w$ becomes negative (with a positive vanishing imaginary part)
  \be
  w = - \frac{\sqrt{s}-\sqrt{s-4m^2}}{\sqrt{s}+\sqrt{s-4m^2}} 
  = - w' + i0^+ \, , 
  \ee
  and $1>w'>0$ when $4m^2<s<+\infty$.
\end{enumerate}

\subsubsection{Range of values for $z$}

The variable $z$ depends both on $s$ and $t$.  In order to study the
different regimes, we define the following function of $s$
\begin{align}
  t_* \equiv {} &  - m^2 \frac{w}{(1+w)^2}  \nonumber\\
  = {} & - \frac{m^4}{4m^2-s}
         \, .
\end{align}
where the second equality follows from eq.~\eqref{eq:w}. 
We also define the ratio
\begin{align}
  \label{eq:defK}
  K \equiv {} & \frac{t}{t_*}\,,
\end{align}
so that the second of eqs.~\eqref{eq:wz} reads
\begin{align}
  \label{eq:zK}
  K = \frac{(1+z)^2}{z}\,.
\end{align}
We choose the following root of the above equation 
\begin{align}
  \label{eq:z}
  z = {} & \frac{\sqrt{K}-\sqrt{K-4}}{\sqrt{K}+\sqrt{K-4}} \,.    
\end{align}
Note that eq.~\eqref{eq:z} contains square-roots of $K$. Therefore, in
order to compute $z$ when $K<4$, we have to keep track of the vanishing
imaginary parts of the quantities entering~eq.~\eqref{eq:defK}.
Region by region in the $(s,t)$ plane, the correct sign of the
vanishing imaginary part (if present) is determined by the Feynman
prescription on $s,t,u$, i.e.\ $s+i0^+$ when $s>0$, and likewise for $t$ and $u$.

Depending on the value of $K$, we distinguish three cases (here we keep the
prescription for the vanishing imaginary part of $K$ arbitrary):

\begin{enumerate}
\item $K>4$

  All the square roots in eq.~\eqref{eq:z} are real, so $z$ is real
  with $0<z<1$.

\item $0<K<4$

  For a given prescription $K\pm i0^+$, one obtains from
  eq.~\eqref{eq:z}
  \begin{align}
    z =\frac{\sqrt{K} \mp i\sqrt{4-K}}{\sqrt{K} \pm i\sqrt{4-K}}  \, ,
  \end{align}
  which is solved by
  \begin{align}
    \label{eq:zphase}
    z = e^{\mp i \psi} \,, \qquad \psi = 2 \arctan{\sqrt{\frac{4-K}{K}}} \,, \qquad 0<\psi<\pi \,.
  \end{align}

\item $K<0$

  For a given prescription $K\pm i0^+$, one obtains from
  eq.~\eqref{eq:z}
  \begin{align}
    \label{eq:zprime}
    z = {} & \frac{\sqrt{-|K| \pm i0^+}-\sqrt{-|K|-4 \pm i0^+}}{{\sqrt{-|K| \pm i0^+}+\sqrt{-|K|-4 \pm i0^+}}} \nonumber\\
    = {} & \frac{\sqrt{|K|}-\sqrt{|K|+4}}{\sqrt{|K|}+\sqrt{|K|+4}} \mp i0^+ \nonumber\\
    \equiv {} & -z' \mp i0^+\,,    
  \end{align}
  with $0<z'<1$.

\end{enumerate}
Note that, since $K$ is a function of $s$ and $t$, each case can arise
from multiple regions in the $(s,t)$ plane.  In
table~\ref{tab:zregions} we summarize the solution for $z$ in the
different regions of the $(s,t)$ plane, by displaying also the
appropriate sign for the $i0^+$ prescription (if a vanishing imaginary
part is present).

\begin{table}
  \centering
  \begin{tabular}{c|ccccc}
    \toprule
    & $t<-4|t_*|$ & $-4|t_*|<t<0$ & $0<t<|t_*|$ & $|t_*|<t<4|t_*|$ & $t>|4t_*|$ \\ 
    \midrule
    $s<0$  & $z$ & $ e^{-i\psi}$ & $-z'+i0^+$ & $-z'+i0^+$ & $-z'+i0^+$ \\
    $0<s<4m^2$  & $z$ & $ \boldsymbol{e^{i\psi}}$ & $-z'+i0^+$ & $-z'-i0^+$ &  $-z'-i0^+$ \\
    $s>4m^2$  & $\boldsymbol{-z'+i0^+}$ & $\boldsymbol{-z'+i0^+}$ & $ e^{-i\psi}$ & $ e^{-i\psi}$ & $z$ \\
    \bottomrule
  \end{tabular}
  \caption{We show the solution for $z$ in each region of the $(s,t)$
    plane, as in
    eqs.~\eqref{eq:z},~\eqref{eq:zphase}, and~\eqref{eq:zprime}. The boldface
    entries are the solutions in the regions that contain a part of
    the physical $s$-channel scattering region, $s>0$ with $-s<t<0$.}
  \label{tab:zregions}
\end{table}


\section[Two-Loop dlog-forms]{Two-Loop $\dlog$-forms}

In this appendix we give explicitly the coefficient matrices of the
$\dlog$-forms, eq.~\eqref{dlog-form}, for the one-mass and the
two-mass two-loop MIs, discussed respectively in
sections~\ref{sec:1mMIs} and~\ref{sec:2mMIs}.

\subsection{One-mass}
\label{dlog-1M2L}

For the one-mass case at the two-loop order, the $\dlog$-form is
\begin{align}
\dA = {} & \MM_1 \, \dlog(1+x) + \MM_2 \, \dlog(x) + \MM_3 \, \dlog(y) \nn
  & + \MM_4 \, \dlog(1-y)+ \MM_5 \, \dlog(x+y)+ \MM_6 \, \dlog(x+y+xy) 
\end{align}
with 

\begin{align}
\MM_1 = \scalemath{0.55}{
\left(
\begin{array}{ccccccccccccccccccccccccccccccc}
 \pminus 0 & \pminus 0 & \pminus 0 & \pminus 0 & \pminus 0 & \pminus 0 & \pminus 0 & \pminus 0 & \pminus 0 & \pminus 0 & \pminus 0 & \pminus 0 & \pminus 0 & \pminus 0 & \pminus 0 & \pminus 0 & \pminus 0 & \pminus 0 & \pminus 0 & \pminus 0 & \pminus 0 & \pminus 0 & \pminus 0 & \pminus 0 & \pminus 0 & \pminus 0 & \pminus 0 & \pminus 0 & \pminus 0 & \pminus 0 & \pminus 0 \\
 \pminus 0 & \pminus 1 & \pminus 0 & \pminus 0 & \pminus 0 & \pminus 0 & \pminus 0 & \pminus 0 & \pminus 0 & \pminus 0 & \pminus 0 & \pminus 0 & \pminus 0 & \pminus 0 & \pminus 0 & \pminus 0 & \pminus 0 & \pminus 0 & \pminus 0 & \pminus 0 & \pminus 0 & \pminus 0 & \pminus 0 & \pminus 0 & \pminus 0 & \pminus 0 & \pminus 0 & \pminus 0 & \pminus 0 & \pminus 0 & \pminus 0 \\
 \pminus 0 & \pminus 4 & \pminus 0 & \pminus 0 & \pminus 0 & \pminus 0 & \pminus 0 & \pminus 0 & \pminus 0 & \pminus 0 & \pminus 0 & \pminus 0 & \pminus 0 & \pminus 0 & \pminus 0 & \pminus 0 & \pminus 0 & \pminus 0 & \pminus 0 & \pminus 0 & \pminus 0 & \pminus 0 & \pminus 0 & \pminus 0 & \pminus 0 & \pminus 0 & \pminus 0 & \pminus 0 & \pminus 0 & \pminus 0 & \pminus 0 \\
 \pminus 0 & \pminus 0 & \pminus 0 & \minus  1 & \pminus 0 & \pminus 0 & \pminus 0 & \pminus 0 & \pminus 0 & \pminus 0 & \pminus 0 & \pminus 0 & \pminus 0 & \pminus 0 & \pminus 0 & \pminus 0 & \pminus 0 & \pminus 0 & \pminus 0 & \pminus 0 & \pminus 0 & \pminus 0 & \pminus 0 & \pminus 0 & \pminus 0 & \pminus 0 & \pminus 0 & \pminus 0 & \pminus 0 & \pminus 0 & \pminus 0 \\
 \pminus 0 & \pminus 0 & \pminus 0 & \pminus 0 & \minus  2 & \pminus 0 & \pminus 0 & \pminus 0 & \pminus 0 & \pminus 0 & \pminus 0 & \pminus 0 & \pminus 0 & \pminus 0 & \pminus 0 & \pminus 0 & \pminus 0 & \pminus 0 & \pminus 0 & \pminus 0 & \pminus 0 & \pminus 0 & \pminus 0 & \pminus 0 & \pminus 0 & \pminus 0 & \pminus 0 & \pminus 0 & \pminus 0 & \pminus 0 & \pminus 0 \\
\end{array}
\right)
  }\,,
\end{align}

\begin{align}
\MM_2 = \scalemath{0.55}{
\left(
\begin{array}{ccccccccccccccccccccccccccccccc}
 \pminus 0 & \pminus 0 & \pminus 0 & \pminus 0 & \pminus 0 & \pminus 0 & \pminus 0 & \pminus 0 & \pminus 0 & \pminus 0 & \pminus 0 & \pminus 0 & \pminus 0 & \pminus 0 & \pminus 0 & \pminus 0 & \pminus 0 & \pminus 0 & \pminus 0 & \pminus 0 & \pminus 0 & \pminus 0 & \pminus 0 & \pminus 0 & \pminus 0 & \pminus 0 & \pminus 0 & \pminus 0 & \pminus 0 & \pminus 0 & \pminus 0 \\
 \pminus 0 & \minus  2 & \minus  1 & \pminus 0 & \pminus 0 & \pminus 0 & \pminus 0 & \pminus 0 & \pminus 0 & \pminus 0 & \pminus 0 & \pminus 0 & \pminus 0 & \pminus 0 & \pminus 0 & \pminus 0 & \pminus 0 & \pminus 0 & \pminus 0 & \pminus 0 & \pminus 0 & \pminus 0 & \pminus 0 & \pminus 0 & \pminus 0 & \pminus 0 & \pminus 0 & \pminus 0 & \pminus 0 & \pminus 0 & \pminus 0 \\
 \pminus 0 & \minus  4 & \minus  2 & \pminus 0 & \pminus 0 & \pminus 0 & \pminus 0 & \pminus 0 & \pminus 0 & \pminus 0 & \pminus 0 & \pminus 0 & \pminus 0 & \pminus 0 & \pminus 0 & \pminus 0 & \pminus 0 & \pminus 0 & \pminus 0 & \pminus 0 & \pminus 0 & \pminus 0 & \pminus 0 & \pminus 0 & \pminus 0 & \pminus 0 & \pminus 0 & \pminus 0 & \pminus 0 & \pminus 0 & \pminus 0 \\
 \pminus 0 & \pminus 0 & \pminus 0 & \pminus 0 & \pminus 0 & \pminus 0 & \pminus 0 & \pminus 0 & \pminus 0 & \pminus 0 & \pminus 0 & \pminus 0 & \pminus 0 & \pminus 0 & \pminus 0 & \pminus 0 & \pminus 0 & \pminus 0 & \pminus 0 & \pminus 0 & \pminus 0 & \pminus 0 & \pminus 0 & \pminus 0 & \pminus 0 & \pminus 0 & \pminus 0 & \pminus 0 & \pminus 0 & \pminus 0 & \pminus 0 \\
 \pminus 0 & \pminus 0 & \pminus 0 & \pminus 0 & \pminus 0 & \pminus 0 & \pminus 0 & \pminus 0 & \pminus 0 & \pminus 0 & \pminus 0 & \pminus 0 & \pminus 0 & \pminus 0 & \pminus 0 & \pminus 0 & \pminus 0 & \pminus 0 & \pminus 0 & \pminus 0 & \pminus 0 & \pminus 0 & \pminus 0 & \pminus 0 & \pminus 0 & \pminus 0 & \pminus 0 & \pminus 0 & \pminus 0 & \pminus 0 & \pminus 0 \\
\end{array}
\right)
  }\,,
\end{align}

\begin{align}
\MM_3 = \scalemath{0.55}{
\left(
\begin{array}{ccccccccccccccccccccccccccccccc}
 \pminus 0 & \pminus 0 & \pminus 0 & \pminus 0 & \pminus 0 & \pminus 0 & \pminus 0 & \pminus 0 & \pminus 0 & \pminus 0 & \pminus 0 & \pminus 0 & \pminus 0 & \pminus 0 & \pminus 0 & \pminus 0 & \pminus 0 & \pminus 0 & \pminus 0 & \pminus 0 & \pminus 0 & \pminus 0 & \pminus 0 & \pminus 0 & \pminus 0 & \pminus 0 & \pminus 0 & \pminus 0 & \pminus 0 & \pminus 0 & \pminus 0 \\
 \pminus 0 & \pminus 0 & \pminus 0 & \pminus 0 & \pminus 0 & \pminus 0 & \pminus 0 & \pminus 0 & \pminus 0 & \pminus 0 & \pminus 0 & \pminus 0 & \pminus 0 & \pminus 0 & \pminus 0 & \pminus 0 & \pminus 0 & \pminus 0 & \pminus 0 & \pminus 0 & \pminus 0 & \pminus 0 & \pminus 0 & \pminus 0 & \pminus 0 & \pminus 0 & \pminus 0 & \pminus 0 & \pminus 0 & \pminus 0 & \pminus 0 \\
 \pminus 0 & \pminus 0 & \pminus 0 & \pminus 0 & \pminus 0 & \pminus 0 & \pminus 0 & \pminus 0 & \pminus 0 & \pminus 0 & \pminus 0 & \pminus 0 & \pminus 0 & \pminus 0 & \pminus 0 & \pminus 0 & \pminus 0 & \pminus 0 & \pminus 0 & \pminus 0 & \pminus 0 & \pminus 0 & \pminus 0 & \pminus 0 & \pminus 0 & \pminus 0 & \pminus 0 & \pminus 0 & \pminus 0 & \pminus 0 & \pminus 0 \\
 \pminus 0 & \pminus 0 & \pminus 0 & \pminus 0 & \pminus 0 & \pminus 0 & \pminus 0 & \pminus 0 & \pminus 0 & \pminus 0 & \pminus 0 & \pminus 0 & \pminus 0 & \pminus 0 & \pminus 0 & \pminus 0 & \pminus 0 & \pminus 0 & \pminus 0 & \pminus 0 & \pminus 0 & \pminus 0 & \pminus 0 & \pminus 0 & \pminus 0 & \pminus 0 & \pminus 0 & \pminus 0 & \pminus 0 & \pminus 0 & \pminus 0 \\
 \pminus 0 & \pminus 0 & \pminus 0 & \pminus 0 & \pminus 0 & \pminus 0 & \pminus 0 & \pminus 0 & \pminus 0 & \pminus 0 & \pminus 0 & \pminus 0 & \pminus 0 & \pminus 0 & \pminus 0 & \pminus 0 & \pminus 0 & \pminus 0 & \pminus 0 & \pminus 0 & \pminus 0 & \pminus 0 & \pminus 0 & \pminus 0 & \pminus 0 & \pminus 0 & \pminus 0 & \pminus 0 & \pminus 0 & \pminus 0 & \pminus 0 \\
\end{array}
\right)
  }\,,
\end{align}

\begin{align}
\MM_4 = \scalemath{0.55}{
\left(
\begin{array}{ccccccccccccccccccccccccccccccc}
 \pminus 0 & \pminus 0 & \pminus 0 & \pminus 0 & \pminus 0 & \pminus 0 & \pminus 0 & \pminus 0 & \pminus 0 & \pminus 0 & \pminus 0 & \pminus 0 & \pminus 0 & \pminus 0 & \pminus 0 & \pminus 0 & \pminus 0 & \pminus 0 & \pminus 0 & \pminus 0 & \pminus 0 & \pminus 0 & \pminus 0 & \pminus 0 & \pminus 0 & \pminus 0 & \pminus 0 & \pminus 0 & \pminus 0 & \pminus 0 & \pminus 0 \\
 \pminus 0 & \pminus 0 & \pminus 0 & \pminus 0 & \pminus 0 & \pminus 0 & \pminus 0 & \pminus 0 & \pminus 0 & \pminus 0 & \pminus 0 & \pminus 0 & \pminus 0 & \pminus 0 & \pminus 0 & \pminus 0 & \pminus 0 & \pminus 0 & \pminus 0 & \pminus 0 & \pminus 0 & \pminus 0 & \pminus 0 & \pminus 0 & \pminus 0 & \pminus 0 & \pminus 0 & \pminus 0 & \pminus 0 & \pminus 0 & \pminus 0 \\
 \pminus 0 & \pminus 0 & \pminus 0 & \pminus 0 & \pminus 0 & \pminus 0 & \pminus 0 & \pminus 0 & \pminus 0 & \pminus 0 & \pminus 0 & \pminus 0 & \pminus 0 & \pminus 0 & \pminus 0 & \pminus 0 & \pminus 0 & \pminus 0 & \pminus 0 & \pminus 0 & \pminus 0 & \pminus 0 & \pminus 0 & \pminus 0 & \pminus 0 & \pminus 0 & \pminus 0 & \pminus 0 & \pminus 0 & \pminus 0 & \pminus 0 \\
 \pminus 0 & \pminus 0 & \pminus 0 & \pminus 0 & \pminus 0 & \pminus 0 & \pminus 0 & \pminus 0 & \pminus 0 & \pminus 0 & \pminus 0 & \pminus 0 & \pminus 0 & \pminus 0 & \pminus 0 & \pminus 0 & \pminus 0 & \pminus 0 & \pminus 0 & \pminus 0 & \pminus 0 & \pminus 0 & \pminus 0 & \pminus 0 & \pminus 0 & \pminus 0 & \pminus 0 & \pminus 0 & \pminus 0 & \pminus 0 & \pminus 0 \\
 \pminus 0 & \pminus 0 & \pminus 0 & \pminus 0 & \pminus 0 & \pminus 0 & \pminus 0 & \pminus 0 & \pminus 0 & \pminus 0 & \pminus 0 & \pminus 0 & \pminus 0 & \pminus 0 & \pminus 0 & \pminus 0 & \pminus 0 & \pminus 0 & \pminus 0 & \pminus 0 & \pminus 0 & \pminus 0 & \pminus 0 & \pminus 0 & \pminus 0 & \pminus 0 & \pminus 0 & \pminus 0 & \pminus 0 & \pminus 0 & \pminus 0 \\
\end{array}
\right)
  }\,,
\end{align}

\begin{align}
\MM_5 = \scalemath{0.55}{
\left(
\begin{array}{ccccccccccccccccccccccccccccccc}
 \pminus 0 & \pminus 0 & \pminus 0 & \pminus 0 & \pminus 0 & \pminus 0 & \pminus 0 & \pminus 0 & \pminus 0 & \pminus 0 & \pminus 0 & \pminus 0 & \pminus 0 & \pminus 0 & \pminus 0 & \pminus 0 & \pminus 0 & \pminus 0 & \pminus 0 & \pminus 0 & \pminus 0 & \pminus 0 & \pminus 0 & \pminus 0 & \pminus 0 & \pminus 0 & \pminus 0 & \pminus 0 & \pminus 0 & \pminus 0 & \pminus 0 \\
 \pminus 0 & \pminus 0 & \pminus 0 & \pminus 0 & \pminus 0 & \pminus 0 & \pminus 0 & \pminus 0 & \pminus 0 & \pminus 0 & \pminus 0 & \pminus 0 & \pminus 0 & \pminus 0 & \pminus 0 & \pminus 0 & \pminus 0 & \pminus 0 & \pminus 0 & \pminus 0 & \pminus 0 & \pminus 0 & \pminus 0 & \pminus 0 & \pminus 0 & \pminus 0 & \pminus 0 & \pminus 0 & \pminus 0 & \pminus 0 & \pminus 0 \\
 \pminus 0 & \pminus 0 & \pminus 0 & \pminus 0 & \pminus 0 & \pminus 0 & \pminus 0 & \pminus 0 & \pminus 0 & \pminus 0 & \pminus 0 & \pminus 0 & \pminus 0 & \pminus 0 & \pminus 0 & \pminus 0 & \pminus 0 & \pminus 0 & \pminus 0 & \pminus 0 & \pminus 0 & \pminus 0 & \pminus 0 & \pminus 0 & \pminus 0 & \pminus 0 & \pminus 0 & \pminus 0 & \pminus 0 & \pminus 0 & \pminus 0 \\
 \pminus 0 & \pminus 0 & \pminus 0 & \pminus 0 & \pminus 0 & \pminus 0 & \pminus 0 & \pminus 0 & \pminus 0 & \pminus 0 & \pminus 0 & \pminus 0 & \pminus 0 & \pminus 0 & \pminus 0 & \pminus 0 & \pminus 0 & \pminus 0 & \pminus 0 & \pminus 0 & \pminus 0 & \pminus 0 & \pminus 0 & \pminus 0 & \pminus 0 & \pminus 0 & \pminus 0 & \pminus 0 & \pminus 0 & \pminus 0 & \pminus 0 \\
 \pminus 0 & \pminus 0 & \pminus 0 & \pminus 0 & \pminus 0 & \pminus 0 & \pminus 0 & \pminus 0 & \pminus 0 & \pminus 0 & \pminus 0 & \pminus 0 & \pminus 0 & \pminus 0 & \pminus 0 & \pminus 0 & \pminus 0 & \pminus 0 & \pminus 0 & \pminus 0 & \pminus 0 & \pminus 0 & \pminus 0 & \pminus 0 & \pminus 0 & \pminus 0 & \pminus 0 & \pminus 0 & \pminus 0 & \pminus 0 & \pminus 0 \\
\end{array}
\right)
  }\,,
\end{align}

\begin{align}
\MM_6 = \scalemath{0.55}{
\left(
\begin{array}{ccccccccccccccccccccccccccccccc}
 \pminus 0 & \pminus 0 & \pminus 0 & \pminus 0 & \pminus 0 & \pminus 0 & \pminus 0 & \pminus 0 & \pminus 0 & \pminus 0 & \pminus 0 & \pminus 0 & \pminus 0 & \pminus 0 & \pminus 0 & \pminus 0 & \pminus 0 & \pminus 0 & \pminus 0 & \pminus 0 & \pminus 0 & \pminus 0 & \pminus 0 & \pminus 0 & \pminus 0 & \pminus 0 & \pminus 0 & \pminus 0 & \pminus 0 & \pminus 0 & \pminus 0 \\
 \pminus 0 & \pminus 0 & \pminus 0 & \pminus 0 & \pminus 0 & \pminus 0 & \pminus 0 & \pminus 0 & \pminus 0 & \pminus 0 & \pminus 0 & \pminus 0 & \pminus 0 & \pminus 0 & \pminus 0 & \pminus 0 & \pminus 0 & \pminus 0 & \pminus 0 & \pminus 0 & \pminus 0 & \pminus 0 & \pminus 0 & \pminus 0 & \pminus 0 & \pminus 0 & \pminus 0 & \pminus 0 & \pminus 0 & \pminus 0 & \pminus 0 \\
 \pminus 0 & \pminus 0 & \pminus 0 & \pminus 0 & \pminus 0 & \pminus 0 & \pminus 0 & \pminus 0 & \pminus 0 & \pminus 0 & \pminus 0 & \pminus 0 & \pminus 0 & \pminus 0 & \pminus 0 & \pminus 0 & \pminus 0 & \pminus 0 & \pminus 0 & \pminus 0 & \pminus 0 & \pminus 0 & \pminus 0 & \pminus 0 & \pminus 0 & \pminus 0 & \pminus 0 & \pminus 0 & \pminus 0 & \pminus 0 & \pminus 0 \\
 \pminus 0 & \pminus 0 & \pminus 0 & \pminus 0 & \pminus 0 & \pminus 0 & \pminus 0 & \pminus 0 & \pminus 0 & \pminus 0 & \pminus 0 & \pminus 0 & \pminus 0 & \pminus 0 & \pminus 0 & \pminus 0 & \pminus 0 & \pminus 0 & \pminus 0 & \pminus 0 & \pminus 0 & \pminus 0 & \pminus 0 & \pminus 0 & \pminus 0 & \pminus 0 & \pminus 0 & \pminus 0 & \pminus 0 & \pminus 0 & \pminus 0 \\
 \pminus 0 & \pminus 0 & \pminus 0 & \pminus 0 & \pminus 0 & \pminus 0 & \pminus 0 & \pminus 0 & \pminus 0 & \pminus 0 & \pminus 0 & \pminus 0 & \pminus 0 & \pminus 0 & \pminus 0 & \pminus 0 & \pminus 0 & \pminus 0 & \pminus 0 & \pminus 0 & \pminus 0 & \pminus 0 & \pminus 0 & \pminus 0 & \pminus 0 & \pminus 0 & \pminus 0 & \pminus 0 & \pminus 0 & \pminus 0 & \pminus 0 \\
\end{array}
\right)
  }\,,
\end{align}

\subsection{Two-mass}

For the two-mass case, at the two-loop order, the $\dlog$-form is
\begin{align}
\dA = {} & \MM_1 \, \dlog(z) + \MM_2 \, \dlog(1+z) + \MM_3 \, \dlog(1-z)+ \MM_4 \, \dlog(w)  \nn
& + \MM_5 \, \dlog(1+w) +\MM_6 \, \dlog(1-w)+\MM_7 \, \dlog(1-w+w^2) \nn
& + \MM_8 \, \dlog(1-w \, z) + \MM_9 \, \dlog(z-w) + \MM_{10} \, \dlog(1+w^2 \, z)\nn
& + \MM_{11} \, \dlog(w^2+z)+ \MM_{12} \, \dlog((1+z)^4 w^3 +(1-w)^2  \, \kappa^2_+(w,z))  \nn
& + \MM_{13} \, \dlog( (1+w)\sqrt{\rho} + (1-w)\, \kappa_-(w,z) ) \nn
& + \MM_{14} \, \dlog( (1+w)\sqrt{\rho} - (1-w)\, \kappa_-(w,z) ) \nn
& + \MM_{15} \, \dlog( (1+w)\sqrt{\rho} + (1-w)\, \kappa_+(w,z) ) \nn
& + \MM_{16} \, \dlog\left( \frac{c_1+c_2 \, \sqrt{\rho}}{c_3+c_4 \, \sqrt{\rho}} \right)\nn
& + \MM_{17} \, \dlog( 2(1-w)^2 w z^2 + \kappa_-^2(-w,z) + \left(z+w \right) \left(1+w  z\right) \sqrt{\rho} )  
\end{align}
where we used the abbreviations introduced below eq.~\eqref{def:2M2Lalphabet}. The
coefficient matrices are

\begin{align}
\MM_1 = \scalemath{0.55}{
\left(
\begin{array}{cccccccccccccccccccccccccccccccccccc}
 \pminus 0 & \pminus 0 & \pminus 0 & \pminus 0 & \pminus 0 & \pminus 0 & \pminus 0 & \pminus 0 & \pminus 0 & \pminus 0 & \pminus 0 & \pminus 0 & \pminus 0 & \pminus 0 & \pminus 0 & \pminus 0 & \pminus 0 & \pminus 0 & \pminus 0 & \pminus 0 & \pminus 0 & \pminus 0 & \pminus 0 & \pminus 0 & \pminus 0 & \pminus 0 & \pminus 0 & \pminus 0 & \pminus 0 & \pminus 0 & \pminus 0 & \pminus 0 & \pminus 0 & \pminus 0 & \pminus 0 & \pminus 0 \\
 \pminus 0 & \pminus 0 & \pminus 0 & \pminus 0 & \pminus 0 & \pminus 0 & \pminus 0 & \pminus 0 & \pminus 0 & \pminus 0 & \pminus 0 & \pminus 0 & \pminus 0 & \pminus 0 & \pminus 0 & \pminus 0 & \pminus 0 & \pminus 0 & \pminus 0 & \pminus 0 & \pminus 0 & \pminus 0 & \pminus 0 & \pminus 0 & \pminus 0 & \pminus 0 & \pminus 0 & \pminus 0 & \pminus 0 & \pminus 0 & \pminus 0 & \pminus 0 & \pminus 0 & \pminus 0 & \pminus 0 & \pminus 0 \\
 \pminus 0 & \pminus 0 & \pminus 0 & \pminus 0 & \pminus 0 & \pminus 0 & \pminus 0 & \pminus 0 & \pminus 0 & \pminus 0 & \pminus 0 & \pminus 0 & \pminus 0 & \pminus 0 & \pminus 0 & \pminus 0 & \pminus 0 & \pminus 0 & \pminus 0 & \pminus 0 & \pminus 0 & \pminus 0 & \pminus 0 & \pminus 0 & \pminus 0 & \pminus 0 & \pminus 0 & \pminus 0 & \pminus 0 & \pminus 0 & \pminus 0 & \pminus 0 & \pminus 0 & \pminus 0 & \pminus 0 & \pminus 0 \\
 \pminus 0 & \pminus 0 & \pminus 0 & \pminus 0 & \pminus 0 & \pminus 0 & \pminus 0 & \pminus 0 & \pminus 0 & \pminus 0 & \pminus 0 & \pminus 0 & \pminus 0 & \pminus 0 & \pminus 0 & \pminus 0 & \pminus 0 & \pminus 0 & \pminus 0 & \pminus 0 & \pminus 0 & \pminus 0 & \pminus 0 & \pminus 0 & \pminus 0 & \pminus 0 & \pminus 0 & \pminus 0 & \pminus 0 & \pminus 0 & \pminus 0 & \pminus 0 & \pminus 0 & \pminus 0 & \pminus 0 & \pminus 0 \\
\end{array}
\right)
  }\,,
\end{align}

\begin{align}
\MM_2 = \scalemath{0.55}{
\left(
\begin{array}{cccccccccccccccccccccccccccccccccccc}
 \pminus 0 & \pminus 0 & \pminus 0 & \pminus 0 & \pminus 0 & \pminus 0 & \pminus 0 & \pminus 0 & \pminus 0 & \pminus 0 & \pminus 0 & \pminus 0 & \pminus 0 & \pminus 0 & \pminus 0 & \pminus 0 & \pminus 0 & \pminus 0 & \pminus 0 & \pminus 0 & \pminus 0 & \pminus 0 & \pminus 0 & \pminus 0 & \pminus 0 & \pminus 0 & \pminus 0 & \pminus 0 & \pminus 0 & \pminus 0 & \pminus 0 & \pminus 0 & \pminus 0 & \pminus 0 & \pminus 0 & \pminus 0 \\
 \pminus 0 & \pminus 0 & \pminus 0 & \pminus 0 & \pminus 0 & \pminus 0 & \pminus 0 & \pminus 0 & \pminus 0 & \pminus 0 & \pminus 0 & \pminus 0 & \pminus 0 & \pminus 0 & \pminus 0 & \pminus 0 & \pminus 0 & \pminus 0 & \pminus 0 & \pminus 0 & \pminus 0 & \pminus 0 & \pminus 0 & \pminus 0 & \pminus 0 & \pminus 0 & \pminus 0 & \pminus 0 & \pminus 0 & \pminus 0 & \pminus 0 & \pminus 0 & \pminus 0 & \pminus 0 & \pminus 0 & \pminus 0 \\
 \pminus 0 & \pminus 0 & \pminus 0 & \pminus 0 & \pminus 0 & \pminus 0 & \pminus 0 & \pminus 0 & \pminus 0 & \pminus 0 & \pminus 0 & \pminus 0 & \pminus 0 & \pminus 0 & \pminus 0 & \pminus 0 & \pminus 0 & \pminus 0 & \pminus 0 & \pminus 0 & \pminus 0 & \pminus 0 & \pminus 0 & \pminus 0 & \pminus 0 & \pminus 0 & \pminus 0 & \pminus 0 & \pminus 0 & \pminus 0 & \pminus 0 & \pminus 0 & \pminus 0 & \pminus 0 & \pminus 0 & \pminus 0 \\
 \pminus 0 & \pminus 0 & \pminus 0 & \pminus 0 & \pminus 0 & \pminus 0 & \pminus 0 & \pminus 0 & \pminus 0 & \pminus 0 & \pminus 0 & \pminus 0 & \pminus 0 & \pminus 0 & \pminus 0 & \pminus 0 & \pminus 0 & \pminus 0 & \pminus 0 & \pminus 0 & \pminus 0 & \pminus 0 & \pminus 0 & \pminus 0 & \pminus 0 & \pminus 0 & \pminus 0 & \pminus 0 & \pminus 0 & \pminus 0 & \pminus 0 & \pminus 0 & \pminus 0 & \pminus 0 & \pminus 0 & \pminus 0 \\
\end{array}
\right)
  }\,,
\end{align}

\begin{align}
\MM_3 = \scalemath{0.55}{
\left(
\begin{array}{cccccccccccccccccccccccccccccccccccc}
 \pminus 0 & \pminus 0 & \pminus 0 & \pminus 0 & \pminus 0 & \pminus 0 & \pminus 0 & \pminus 0 & \pminus 0 & \pminus 0 & \pminus 0 & \pminus 0 & \pminus 0 & \pminus 0 & \pminus 0 & \pminus 0 & \pminus 0 & \pminus 0 & \pminus 0 & \pminus 0 & \pminus 0 & \pminus 0 & \pminus 0 & \pminus 0 & \pminus 0 & \pminus 0 & \pminus 0 & \pminus 0 & \pminus 0 & \pminus 0 & \pminus 0 & \pminus 0 & \pminus 0 & \pminus 0 & \pminus 0 & \pminus 0 \\
 \pminus 0 & \pminus 0 & \pminus 0 & \pminus 0 & \pminus 0 & \pminus 0 & \pminus 0 & \pminus 0 & \pminus 0 & \pminus 0 & \pminus 0 & \pminus 0 & \pminus 0 & \pminus 0 & \pminus 0 & \pminus 0 & \pminus 0 & \pminus 0 & \pminus 0 & \pminus 0 & \pminus 0 & \pminus 0 & \pminus 0 & \pminus 0 & \pminus 0 & \pminus 0 & \pminus 0 & \pminus 0 & \pminus 0 & \pminus 0 & \pminus 0 & \pminus 0 & \pminus 0 & \pminus 0 & \pminus 0 & \pminus 0 \\
 \pminus 0 & \pminus 0 & \pminus 0 & \pminus 0 & \pminus 0 & \pminus 0 & \pminus 0 & \pminus 0 & \pminus 0 & \pminus 0 & \pminus 0 & \pminus 0 & \pminus 0 & \pminus 0 & \pminus 0 & \pminus 0 & \pminus 0 & \pminus 0 & \pminus 0 & \pminus 0 & \pminus 0 & \pminus 0 & \pminus 0 & \pminus 0 & \pminus 0 & \pminus 0 & \pminus 0 & \pminus 0 & \pminus 0 & \pminus 0 & \pminus 0 & \pminus 0 & \pminus 0 & \pminus 0 & \pminus 0 & \pminus 0 \\
 \pminus 0 & \pminus 0 & \pminus 0 & \pminus 0 & \pminus 0 & \pminus 0 & \pminus 0 & \pminus 0 & \pminus 0 & \pminus 0 & \pminus 0 & \pminus 0 & \pminus 0 & \pminus 0 & \pminus 0 & \pminus 0 & \pminus 0 & \pminus 0 & \pminus 0 & \pminus 0 & \pminus 0 & \pminus 0 & \pminus 0 & \pminus 0 & \pminus 0 & \pminus 0 & \pminus 0 & \pminus 0 & \pminus 0 & \pminus 0 & \pminus 0 & \pminus 0 & \pminus 0 & \pminus 0 & \pminus 0 & \pminus 0 \\
\end{array}
\right)
  }\,,
\end{align}

\begin{align}
\MM_4 = \scalemath{0.55}{
\left(
\begin{array}{cccccccccccccccccccccccccccccccccccc}
 \pminus 0 & \pminus 0 & \pminus 0 & \pminus 0 & \pminus 0 & \pminus 0 & \pminus 0 & \pminus 0 & \pminus 0 & \pminus 0 & \pminus 0 & \pminus 0 & \pminus 0 & \pminus 0 & \pminus 0 & \pminus 0 & \pminus 0 & \pminus 0 & \pminus 0 & \pminus 0 & \pminus 0 & \pminus 0 & \pminus 0 & \pminus 0 & \pminus 0 & \pminus 0 & \pminus 0 & \pminus 0 & \pminus 0 & \pminus 0 & \pminus 0 & \pminus 0 & \pminus 0 & \pminus 0 & \pminus 0 & \pminus 0 \\
 \pminus 0 & \pminus 1 & \pminus 1 & \pminus 0 & \pminus 0 & \pminus 0 & \pminus 0 & \pminus 0 & \pminus 0 & \pminus 0 & \pminus 0 & \pminus 0 & \pminus 0 & \pminus 0 & \pminus 0 & \pminus 0 & \pminus 0 & \pminus 0 & \pminus 0 & \pminus 0 & \pminus 0 & \pminus 0 & \pminus 0 & \pminus 0 & \pminus 0 & \pminus 0 & \pminus 0 & \pminus 0 & \pminus 0 & \pminus 0 & \pminus 0 & \pminus 0 & \pminus 0 & \pminus 0 & \pminus 0 & \pminus 0 \\
 \pminus 0 & \pminus 0 & \pminus 2 & \pminus 0 & \pminus 0 & \pminus 0 & \pminus 0 & \pminus 0 & \pminus 0 & \pminus 0 & \pminus 0 & \pminus 0 & \pminus 0 & \pminus 0 & \pminus 0 & \pminus 0 & \pminus 0 & \pminus 0 & \pminus 0 & \pminus 0 & \pminus 0 & \pminus 0 & \pminus 0 & \pminus 0 & \pminus 0 & \pminus 0 & \pminus 0 & \pminus 0 & \pminus 0 & \pminus 0 & \pminus 0 & \pminus 0 & \pminus 0 & \pminus 0 & \pminus 0 & \pminus 0 \\
 \pminus 0 & \pminus 0 & \pminus 0 & \pminus 1 & \pminus 0 & \pminus 0 & \pminus 0 & \pminus 0 & \pminus 0 & \pminus 0 & \pminus 0 & \pminus 0 & \pminus 0 & \pminus 0 & \pminus 0 & \pminus 0 & \pminus 0 & \pminus 0 & \pminus 0 & \pminus 0 & \pminus 0 & \pminus 0 & \pminus 0 & \pminus 0 & \pminus 0 & \pminus 0 & \pminus 0 & \pminus 0 & \pminus 0 & \pminus 0 & \pminus 0 & \pminus 0 & \pminus 0 & \pminus 0 & \pminus 0 & \pminus 0 \\
\end{array}
\right)
  }\,,
\end{align}

\begin{align}
\MM_5 = \scalemath{0.55}{
\left(
\begin{array}{cccccccccccccccccccccccccccccccccccc}
 \pminus 0 & \pminus 0 & \pminus 0 & \pminus 0 & \pminus 0 & \pminus 0 & \pminus 0 & \pminus 0 & \pminus 0 & \pminus 0 & \pminus 0 & \pminus 0 & \pminus 0 & \pminus 0 & \pminus 0 & \pminus 0 & \pminus 0 & \pminus 0 & \pminus 0 & \pminus 0 & \pminus 0 & \pminus 0 & \pminus 0 & \pminus 0 & \pminus 0 & \pminus 0 & \pminus 0 & \pminus 0 & \pminus 0 & \pminus 0 & \pminus 0 & \pminus 0 & \pminus 0 & \pminus 0 & \pminus 0 & \pminus 0 \\
 \pminus 0 & \pminus 0 & \pminus 0 & \pminus 0 & \pminus 0 & \pminus 0 & \pminus 0 & \pminus 0 & \pminus 0 & \pminus 0 & \pminus 0 & \pminus 0 & \pminus 0 & \pminus 0 & \pminus 0 & \pminus 0 & \pminus 0 & \pminus 0 & \pminus 0 & \pminus 0 & \pminus 0 & \pminus 0 & \pminus 0 & \pminus 0 & \pminus 0 & \pminus 0 & \pminus 0 & \pminus 0 & \pminus 0 & \pminus 0 & \pminus 0 & \pminus 0 & \pminus 0 & \pminus 0 & \pminus 0 & \pminus 0 \\
 \pminus 0 & \pminus 0 & \pminus 0 & \pminus 0 & \pminus 0 & \pminus 0 & \pminus 0 & \pminus 0 & \pminus 0 & \pminus 0 & \pminus 0 & \pminus 0 & \pminus 0 & \pminus 0 & \pminus 0 & \pminus 0 & \pminus 0 & \pminus 0 & \pminus 0 & \pminus 0 & \pminus 0 & \pminus 0 & \pminus 0 & \pminus 0 & \pminus 0 & \pminus 0 & \pminus 0 & \pminus 0 & \pminus 0 & \pminus 0 & \pminus 0 & \pminus 0 & \pminus 0 & \pminus 0 & \pminus 0 & \pminus 0 \\
 \pminus 0 & \pminus 0 & \pminus 0 & \pminus 0 & \pminus 0 & \pminus 0 & \pminus 0 & \pminus 0 & \pminus 0 & \pminus 0 & \pminus 0 & \pminus 0 & \pminus 0 & \pminus 0 & \pminus 0 & \pminus 0 & \pminus 0 & \pminus 0 & \pminus 0 & \pminus 0 & \pminus 0 & \pminus 0 & \pminus 0 & \pminus 0 & \pminus 0 & \pminus 0 & \pminus 0 & \pminus 0 & \pminus 0 & \pminus 0 & \pminus 0 & \pminus 0 & \pminus 0 & \pminus 0 & \pminus 0 & \pminus 0 \\
\end{array}
\right)
  }\,,
\end{align}

\begin{align}
\MM_6 = \scalemath{0.55}{
\left(
\begin{array}{cccccccccccccccccccccccccccccccccccc}
 \pminus 0 & \pminus 0 & \pminus 0 & \pminus 0 & \pminus 0 & \pminus 0 & \pminus 0 & \pminus 0 & \pminus 0 & \pminus 0 & \pminus 0 & \pminus 0 & \pminus 0 & \pminus 0 & \pminus 0 & \pminus 0 & \pminus 0 & \pminus 0 & \pminus 0 & \pminus 0 & \pminus 0 & \pminus 0 & \pminus 0 & \pminus 0 & \pminus 0 & \pminus 0 & \pminus 0 & \pminus 0 & \pminus 0 & \pminus 0 & \pminus 0 & \pminus 0 & \pminus 0 & \pminus 0 & \pminus 0 & \pminus 0 \\
 \pminus 0 & \pminus 2 & \pminus 0 & \pminus 0 & \pminus 0 & \pminus 0 & \pminus 0 & \pminus 0 & \pminus 0 & \pminus 0 & \pminus 0 & \pminus 0 & \pminus 0 & \pminus 0 & \pminus 0 & \pminus 0 & \pminus 0 & \pminus 0 & \pminus 0 & \pminus 0 & \pminus 0 & \pminus 0 & \pminus 0 & \pminus 0 & \pminus 0 & \pminus 0 & \pminus 0 & \pminus 0 & \pminus 0 & \pminus 0 & \pminus 0 & \pminus 0 & \pminus 0 & \pminus 0 & \pminus 0 & \pminus 0 \\
 \pminus 0 & \pminus 8 & \pminus 0 & \pminus 0 & \pminus 0 & \pminus 0 & \pminus 0 & \pminus 0 & \pminus 0 & \pminus 0 & \pminus 0 & \pminus 0 & \pminus 0 & \pminus 0 & \pminus 0 & \pminus 0 & \pminus 0 & \pminus 0 & \pminus 0 & \pminus 0 & \pminus 0 & \pminus 0 & \pminus 0 & \pminus 0 & \pminus 0 & \pminus 0 & \pminus 0 & \pminus 0 & \pminus 0 & \pminus 0 & \pminus 0 & \pminus 0 & \pminus 0 & \pminus 0 & \pminus 0 & \pminus 0 \\
 \pminus 0 & \pminus 0 & \pminus 0 & \minus  2 & \pminus 0 & \pminus 0 & \pminus 0 & \pminus 0 & \pminus 0 & \pminus 0 & \pminus 0 & \pminus 0 & \pminus 0 & \pminus 0 & \pminus 0 & \pminus 0 & \pminus 0 & \pminus 0 & \pminus 0 & \pminus 0 & \pminus 0 & \pminus 0 & \pminus 0 & \pminus 0 & \pminus 0 & \pminus 0 & \pminus 0 & \pminus 0 & \pminus 0 & \pminus 0 & \pminus 0 & \pminus 0 & \pminus 0 & \pminus 0 & \pminus 0 & \pminus 0 \\
\end{array}
\right)
  }\,,
\end{align}

\begin{align}
\MM_7 = \scalemath{0.55}{
\left(
\begin{array}{cccccccccccccccccccccccccccccccccccc}
 \pminus 0 & \pminus 0 & \pminus 0 & \pminus 0 & \pminus 0 & \pminus 0 & \pminus 0 & \pminus 0 & \pminus 0 & \pminus 0 & \pminus 0 & \pminus 0 & \pminus 0 & \pminus 0 & \pminus 0 & \pminus 0 & \pminus 0 & \pminus 0 & \pminus 0 & \pminus 0 & \pminus 0 & \pminus 0 & \pminus 0 & \pminus 0 & \pminus 0 & \pminus 0 & \pminus 0 & \pminus 0 & \pminus 0 & \pminus 0 & \pminus 0 & \pminus 0 & \pminus 0 & \pminus 0 & \pminus 0 & \pminus 0 \\
 \pminus 0 & \minus  2 & \minus  1 & \pminus 0 & \pminus 0 & \pminus 0 & \pminus 0 & \pminus 0 & \pminus 0 & \pminus 0 & \pminus 0 & \pminus 0 & \pminus 0 & \pminus 0 & \pminus 0 & \pminus 0 & \pminus 0 & \pminus 0 & \pminus 0 & \pminus 0 & \pminus 0 & \pminus 0 & \pminus 0 & \pminus 0 & \pminus 0 & \pminus 0 & \pminus 0 & \pminus 0 & \pminus 0 & \pminus 0 & \pminus 0 & \pminus 0 & \pminus 0 & \pminus 0 & \pminus 0 & \pminus 0 \\
 \pminus 0 & \minus  4 & \minus  2 & \pminus 0 & \pminus 0 & \pminus 0 & \pminus 0 & \pminus 0 & \pminus 0 & \pminus 0 & \pminus 0 & \pminus 0 & \pminus 0 & \pminus 0 & \pminus 0 & \pminus 0 & \pminus 0 & \pminus 0 & \pminus 0 & \pminus 0 & \pminus 0 & \pminus 0 & \pminus 0 & \pminus 0 & \pminus 0 & \pminus 0 & \pminus 0 & \pminus 0 & \pminus 0 & \pminus 0 & \pminus 0 & \pminus 0 & \pminus 0 & \pminus 0 & \pminus 0 & \pminus 0 \\
 \pminus 0 & \pminus 0 & \pminus 0 & \pminus 0 & \pminus 0 & \pminus 0 & \pminus 0 & \pminus 0 & \pminus 0 & \pminus 0 & \pminus 0 & \pminus 0 & \pminus 0 & \pminus 0 & \pminus 0 & \pminus 0 & \pminus 0 & \pminus 0 & \pminus 0 & \pminus 0 & \pminus 0 & \pminus 0 & \pminus 0 & \pminus 0 & \pminus 0 & \pminus 0 & \pminus 0 & \pminus 0 & \pminus 0 & \pminus 0 & \pminus 0 & \pminus 0 & \pminus 0 & \pminus 0 & \pminus 0 & \pminus 0 \\
\end{array}
\right)
  }\,,
\end{align}

\begin{align}
\MM_8 = \scalemath{0.55}{
\left(
\begin{array}{cccccccccccccccccccccccccccccccccccc}
 \pminus 0 & \pminus 0 & \pminus 0 & \pminus 0 & \pminus 0 & \pminus 0 & \pminus 0 & \pminus 0 & \pminus 0 & \pminus 0 & \pminus 0 & \pminus 0 & \pminus 0 & \pminus 0 & \pminus 0 & \pminus 0 & \pminus 0 & \pminus 0 & \pminus 0 & \pminus 0 & \pminus 0 & \pminus 0 & \pminus 0 & \pminus 0 & \pminus 0 & \pminus 0 & \pminus 0 & \pminus 0 & \pminus 0 & \pminus 0 & \pminus 0 & \pminus 0 & \pminus 0 & \pminus 0 & \pminus 0 & \pminus 0 \\
 \pminus 0 & \pminus 0 & \pminus 0 & \pminus 0 & \pminus 0 & \pminus 0 & \pminus 0 & \pminus 0 & \pminus 0 & \pminus 0 & \pminus 0 & \pminus 0 & \pminus 0 & \pminus 0 & \pminus 0 & \pminus 0 & \pminus 0 & \pminus 0 & \pminus 0 & \pminus 0 & \pminus 0 & \pminus 0 & \pminus 0 & \pminus 0 & \pminus 0 & \pminus 0 & \pminus 0 & \pminus 0 & \pminus 0 & \pminus 0 & \pminus 0 & \pminus 0 & \pminus 0 & \pminus 0 & \pminus 0 & \pminus 0 \\
 \pminus 0 & \pminus 0 & \pminus 0 & \pminus 0 & \pminus 0 & \pminus 0 & \pminus 0 & \pminus 0 & \pminus 0 & \pminus 0 & \pminus 0 & \pminus 0 & \pminus 0 & \pminus 0 & \pminus 0 & \pminus 0 & \pminus 0 & \pminus 0 & \pminus 0 & \pminus 0 & \pminus 0 & \pminus 0 & \pminus 0 & \pminus 0 & \pminus 0 & \pminus 0 & \pminus 0 & \pminus 0 & \pminus 0 & \pminus 0 & \pminus 0 & \pminus 0 & \pminus 0 & \pminus 0 & \pminus 0 & \pminus 0 \\
 \pminus 0 & \pminus 0 & \pminus 0 & \pminus 0 & \pminus 0 & \pminus 0 & \pminus 0 & \pminus 0 & \pminus 0 & \pminus 0 & \pminus 0 & \pminus 0 & \pminus 0 & \pminus 0 & \pminus 0 & \pminus 0 & \pminus 0 & \pminus 0 & \pminus 0 & \pminus 0 & \pminus 0 & \pminus 0 & \pminus 0 & \pminus 0 & \pminus 0 & \pminus 0 & \pminus 0 & \pminus 0 & \pminus 0 & \pminus 0 & \pminus 0 & \pminus 0 & \pminus 0 & \pminus 0 & \pminus 0 & \pminus 0 \\
\end{array}
\right)
  }\,,
\end{align}

\begin{align}
\MM_9 = \scalemath{0.55}{
\left(
\begin{array}{cccccccccccccccccccccccccccccccccccc}
 \pminus 0 & \pminus 0 & \pminus 0 & \pminus 0 & \pminus 0 & \pminus 0 & \pminus 0 & \pminus 0 & \pminus 0 & \pminus 0 & \pminus 0 & \pminus 0 & \pminus 0 & \pminus 0 & \pminus 0 & \pminus 0 & \pminus 0 & \pminus 0 & \pminus 0 & \pminus 0 & \pminus 0 & \pminus 0 & \pminus 0 & \pminus 0 & \pminus 0 & \pminus 0 & \pminus 0 & \pminus 0 & \pminus 0 & \pminus 0 & \pminus 0 & \pminus 0 & \pminus 0 & \pminus 0 & \pminus 0 & \pminus 0 \\
 \pminus 0 & \pminus 0 & \pminus 0 & \pminus 0 & \pminus 0 & \pminus 0 & \pminus 0 & \pminus 0 & \pminus 0 & \pminus 0 & \pminus 0 & \pminus 0 & \pminus 0 & \pminus 0 & \pminus 0 & \pminus 0 & \pminus 0 & \pminus 0 & \pminus 0 & \pminus 0 & \pminus 0 & \pminus 0 & \pminus 0 & \pminus 0 & \pminus 0 & \pminus 0 & \pminus 0 & \pminus 0 & \pminus 0 & \pminus 0 & \pminus 0 & \pminus 0 & \pminus 0 & \pminus 0 & \pminus 0 & \pminus 0 \\
 \pminus 0 & \pminus 0 & \pminus 0 & \pminus 0 & \pminus 0 & \pminus 0 & \pminus 0 & \pminus 0 & \pminus 0 & \pminus 0 & \pminus 0 & \pminus 0 & \pminus 0 & \pminus 0 & \pminus 0 & \pminus 0 & \pminus 0 & \pminus 0 & \pminus 0 & \pminus 0 & \pminus 0 & \pminus 0 & \pminus 0 & \pminus 0 & \pminus 0 & \pminus 0 & \pminus 0 & \pminus 0 & \pminus 0 & \pminus 0 & \pminus 0 & \pminus 0 & \pminus 0 & \pminus 0 & \pminus 0 & \pminus 0 \\
 \pminus 0 & \pminus 0 & \pminus 0 & \pminus 0 & \pminus 0 & \pminus 0 & \pminus 0 & \pminus 0 & \pminus 0 & \pminus 0 & \pminus 0 & \pminus 0 & \pminus 0 & \pminus 0 & \pminus 0 & \pminus 0 & \pminus 0 & \pminus 0 & \pminus 0 & \pminus 0 & \pminus 0 & \pminus 0 & \pminus 0 & \pminus 0 & \pminus 0 & \pminus 0 & \pminus 0 & \pminus 0 & \pminus 0 & \pminus 0 & \pminus 0 & \pminus 0 & \pminus 0 & \pminus 0 & \pminus 0 & \pminus 0 \\
\end{array}
\right)
  }\,,
\end{align}

\begin{align}
\MM_{10} = \scalemath{0.55}{
\left(
\begin{array}{cccccccccccccccccccccccccccccccccccc}
 \pminus 0 & \pminus 0 & \pminus 0 & \pminus 0 & \pminus 0 & \pminus 0 & \pminus 0 & \pminus 0 & \pminus 0 & \pminus 0 & \pminus 0 & \pminus 0 & \pminus 0 & \pminus 0 & \pminus 0 & \pminus 0 & \pminus 0 & \pminus 0 & \pminus 0 & \pminus 0 & \pminus 0 & \pminus 0 & \pminus 0 & \pminus 0 & \pminus 0 & \pminus 0 & \pminus 0 & \pminus 0 & \pminus 0 & \pminus 0 & \pminus 0 & \pminus 0 & \pminus 0 & \pminus 0 & \pminus 0 & \pminus 0 \\
 \pminus 0 & \pminus 0 & \pminus 0 & \pminus 0 & \pminus 0 & \pminus 0 & \pminus 0 & \pminus 0 & \pminus 0 & \pminus 0 & \pminus 0 & \pminus 0 & \pminus 0 & \pminus 0 & \pminus 0 & \pminus 0 & \pminus 0 & \pminus 0 & \pminus 0 & \pminus 0 & \pminus 0 & \pminus 0 & \pminus 0 & \pminus 0 & \pminus 0 & \pminus 0 & \pminus 0 & \pminus 0 & \pminus 0 & \pminus 0 & \pminus 0 & \pminus 0 & \pminus 0 & \pminus 0 & \pminus 0 & \pminus 0 \\
 \pminus 0 & \pminus 0 & \pminus 0 & \pminus 0 & \pminus 0 & \pminus 0 & \pminus 0 & \pminus 0 & \pminus 0 & \pminus 0 & \pminus 0 & \pminus 0 & \pminus 0 & \pminus 0 & \pminus 0 & \pminus 0 & \pminus 0 & \pminus 0 & \pminus 0 & \pminus 0 & \pminus 0 & \pminus 0 & \pminus 0 & \pminus 0 & \pminus 0 & \pminus 0 & \pminus 0 & \pminus 0 & \pminus 0 & \pminus 0 & \pminus 0 & \pminus 0 & \pminus 0 & \pminus 0 & \pminus 0 & \pminus 0 \\
 \pminus 0 & \pminus 0 & \pminus 0 & \pminus 0 & \pminus 0 & \pminus 0 & \pminus 0 & \pminus 0 & \pminus 0 & \pminus 0 & \pminus 0 & \pminus 0 & \pminus 0 & \pminus 0 & \pminus 0 & \pminus 0 & \pminus 0 & \pminus 0 & \pminus 0 & \pminus 0 & \pminus 0 & \pminus 0 & \pminus 0 & \pminus 0 & \pminus 0 & \pminus 0 & \pminus 0 & \pminus 0 & \pminus 0 & \pminus 0 & \pminus 0 & \pminus 0 & \pminus 0 & \pminus 0 & \pminus 0 & \pminus 0 \\
\end{array}
\right)
  }\,,
\end{align}

\begin{align}
\MM_{11} = \scalemath{0.55}{
\left(
\begin{array}{cccccccccccccccccccccccccccccccccccc}
 \pminus 0 & \pminus 0 & \pminus 0 & \pminus 0 & \pminus 0 & \pminus 0 & \pminus 0 & \pminus 0 & \pminus 0 & \pminus 0 & \pminus 0 & \pminus 0 & \pminus 0 & \pminus 0 & \pminus 0 & \pminus 0 & \pminus 0 & \pminus 0 & \pminus 0 & \pminus 0 & \pminus 0 & \pminus 0 & \pminus 0 & \pminus 0 & \pminus 0 & \pminus 0 & \pminus 0 & \pminus 0 & \pminus 0 & \pminus 0 & \pminus 0 & \pminus 0 & \pminus 0 & \pminus 0 & \pminus 0 & \pminus 0 \\
 \pminus 0 & \pminus 0 & \pminus 0 & \pminus 0 & \pminus 0 & \pminus 0 & \pminus 0 & \pminus 0 & \pminus 0 & \pminus 0 & \pminus 0 & \pminus 0 & \pminus 0 & \pminus 0 & \pminus 0 & \pminus 0 & \pminus 0 & \pminus 0 & \pminus 0 & \pminus 0 & \pminus 0 & \pminus 0 & \pminus 0 & \pminus 0 & \pminus 0 & \pminus 0 & \pminus 0 & \pminus 0 & \pminus 0 & \pminus 0 & \pminus 0 & \pminus 0 & \pminus 0 & \pminus 0 & \pminus 0 & \pminus 0 \\
 \pminus 0 & \pminus 0 & \pminus 0 & \pminus 0 & \pminus 0 & \pminus 0 & \pminus 0 & \pminus 0 & \pminus 0 & \pminus 0 & \pminus 0 & \pminus 0 & \pminus 0 & \pminus 0 & \pminus 0 & \pminus 0 & \pminus 0 & \pminus 0 & \pminus 0 & \pminus 0 & \pminus 0 & \pminus 0 & \pminus 0 & \pminus 0 & \pminus 0 & \pminus 0 & \pminus 0 & \pminus 0 & \pminus 0 & \pminus 0 & \pminus 0 & \pminus 0 & \pminus 0 & \pminus 0 & \pminus 0 & \pminus 0 \\
 \pminus 0 & \pminus 0 & \pminus 0 & \pminus 0 & \pminus 0 & \pminus 0 & \pminus 0 & \pminus 0 & \pminus 0 & \pminus 0 & \pminus 0 & \pminus 0 & \pminus 0 & \pminus 0 & \pminus 0 & \pminus 0 & \pminus 0 & \pminus 0 & \pminus 0 & \pminus 0 & \pminus 0 & \pminus 0 & \pminus 0 & \pminus 0 & \pminus 0 & \pminus 0 & \pminus 0 & \pminus 0 & \pminus 0 & \pminus 0 & \pminus 0 & \pminus 0 & \pminus 0 & \pminus 0 & \pminus 0 & \pminus 0 \\
\end{array}
\right)
  }\,,
\end{align}


\begin{align}
\MM_{13} = \scalemath{0.55}{
\left(
\begin{array}{cccccccccccccccccccccccccccccccccccc}
 \pminus 0 & \pminus 0 & \pminus 0 & \pminus 0 & \pminus 0 & \pminus 0 & \pminus 0 & \pminus 0 & \pminus 0 & \pminus 0 & \pminus 0 & \pminus 0 & \pminus 0 & \pminus 0 & \pminus 0 & \pminus 0 & \pminus 0 & \pminus 0 & \pminus 0 & \pminus 0 & \pminus 0 & \pminus 0 & \pminus 0 & \pminus 0 & \pminus 0 & \pminus 0 & \pminus 0 & \pminus 0 & \pminus 0 & \pminus 0 & \pminus 0 & \pminus 0 & \pminus 0 & \pminus 0 & \pminus 0 & \pminus 0 \\
 \pminus 0 & \pminus 0 & \pminus 0 & \pminus 0 & \pminus 0 & \pminus 0 & \pminus 0 & \pminus 0 & \pminus 0 & \pminus 0 & \pminus 0 & \pminus 0 & \pminus 0 & \pminus 0 & \pminus 0 & \pminus 0 & \pminus 0 & \pminus 0 & \pminus 0 & \pminus 0 & \pminus 0 & \pminus 0 & \pminus 0 & \pminus 0 & \pminus 0 & \pminus 0 & \pminus 0 & \pminus 0 & \pminus 0 & \pminus 0 & \pminus 0 & \pminus 0 & \pminus 0 & \pminus 0 & \pminus 0 & \pminus 0 \\
 \pminus 0 & \pminus 0 & \pminus 0 & \pminus 0 & \pminus 0 & \pminus 0 & \pminus 0 & \pminus 0 & \pminus 0 & \pminus 0 & \pminus 0 & \pminus 0 & \pminus 0 & \pminus 0 & \pminus 0 & \pminus 0 & \pminus 0 & \pminus 0 & \pminus 0 & \pminus 0 & \pminus 0 & \pminus 0 & \pminus 0 & \pminus 0 & \pminus 0 & \pminus 0 & \pminus 0 & \pminus 0 & \pminus 0 & \pminus 0 & \pminus 0 & \pminus 0 & \pminus 0 & \pminus 0 & \pminus 0 & \pminus 0 \\
 \pminus 0 & \pminus 0 & \pminus 0 & \pminus 0 & \pminus 0 & \pminus 0 & \pminus 0 & \pminus 0 & \pminus 0 & \pminus 0 & \pminus 0 & \pminus 0 & \pminus 0 & \pminus 0 & \pminus 0 & \pminus 0 & \pminus 0 & \pminus 0 & \pminus 0 & \pminus 0 & \pminus 0 & \pminus 0 & \pminus 0 & \pminus 0 & \pminus 0 & \pminus 0 & \pminus 0 & \pminus 0 & \pminus 0 & \pminus 0 & \pminus 0 & \pminus 0 & \pminus 0 & \pminus 0 & \pminus 0 & \pminus 0 \\
\end{array}
\right)
  }\,,
\end{align}

\begin{align}
\MM_{14} = \scalemath{0.55}{
\left(
\begin{array}{cccccccccccccccccccccccccccccccccccc}
 \pminus 0 & \pminus 0 & \pminus 0 & \pminus 0 & \pminus 0 & \pminus 0 & \pminus 0 & \pminus 0 & \pminus 0 & \pminus 0 & \pminus 0 & \pminus 0 & \pminus 0 & \pminus 0 & \pminus 0 & \pminus 0 & \pminus 0 & \pminus 0 & \pminus 0 & \pminus 0 & \pminus 0 & \pminus 0 & \pminus 0 & \pminus 0 & \pminus 0 & \pminus 0 & \pminus 0 & \pminus 0 & \pminus 0 & \pminus 0 & \pminus 0 & \pminus 0 & \pminus 0 & \pminus 0 & \pminus 0 & \pminus 0 \\
 \pminus 0 & \pminus 0 & \pminus 0 & \pminus 0 & \pminus 0 & \pminus 0 & \pminus 0 & \pminus 0 & \pminus 0 & \pminus 0 & \pminus 0 & \pminus 0 & \pminus 0 & \pminus 0 & \pminus 0 & \pminus 0 & \pminus 0 & \pminus 0 & \pminus 0 & \pminus 0 & \pminus 0 & \pminus 0 & \pminus 0 & \pminus 0 & \pminus 0 & \pminus 0 & \pminus 0 & \pminus 0 & \pminus 0 & \pminus 0 & \pminus 0 & \pminus 0 & \pminus 0 & \pminus 0 & \pminus 0 & \pminus 0 \\
 \pminus 0 & \pminus 0 & \pminus 0 & \pminus 0 & \pminus 0 & \pminus 0 & \pminus 0 & \pminus 0 & \pminus 0 & \pminus 0 & \pminus 0 & \pminus 0 & \pminus 0 & \pminus 0 & \pminus 0 & \pminus 0 & \pminus 0 & \pminus 0 & \pminus 0 & \pminus 0 & \pminus 0 & \pminus 0 & \pminus 0 & \pminus 0 & \pminus 0 & \pminus 0 & \pminus 0 & \pminus 0 & \pminus 0 & \pminus 0 & \pminus 0 & \pminus 0 & \pminus 0 & \pminus 0 & \pminus 0 & \pminus 0 \\
 \pminus 0 & \pminus 0 & \pminus 0 & \pminus 0 & \pminus 0 & \pminus 0 & \pminus 0 & \pminus 0 & \pminus 0 & \pminus 0 & \pminus 0 & \pminus 0 & \pminus 0 & \pminus 0 & \pminus 0 & \pminus 0 & \pminus 0 & \pminus 0 & \pminus 0 & \pminus 0 & \pminus 0 & \pminus 0 & \pminus 0 & \pminus 0 & \pminus 0 & \pminus 0 & \pminus 0 & \pminus 0 & \pminus 0 & \pminus 0 & \pminus 0 & \pminus 0 & \pminus 0 & \pminus 0 & \pminus 0 & \pminus 0 \\
\end{array}
\right)
  }\,,
\end{align}

\begin{align}
\MM_{15} = \scalemath{0.55}{
\left(
\begin{array}{cccccccccccccccccccccccccccccccccccc}
 \pminus 0 & \pminus 0 & \pminus 0 & \pminus 0 & \pminus 0 & \pminus 0 & \pminus 0 & \pminus 0 & \pminus 0 & \pminus 0 & \pminus 0 & \pminus 0 & \pminus 0 & \pminus 0 & \pminus 0 & \pminus 0 & \pminus 0 & \pminus 0 & \pminus 0 & \pminus 0 & \pminus 0 & \pminus 0 & \pminus 0 & \pminus 0 & \pminus 0 & \pminus 0 & \pminus 0 & \pminus 0 & \pminus 0 & \pminus 0 & \pminus 0 & \pminus 0 & \pminus 0 & \pminus 0 & \pminus 0 & \pminus 0 \\
 \pminus 0 & \pminus 0 & \pminus 0 & \pminus 0 & \pminus 0 & \pminus 0 & \pminus 0 & \pminus 0 & \pminus 0 & \pminus 0 & \pminus 0 & \pminus 0 & \pminus 0 & \pminus 0 & \pminus 0 & \pminus 0 & \pminus 0 & \pminus 0 & \pminus 0 & \pminus 0 & \pminus 0 & \pminus 0 & \pminus 0 & \pminus 0 & \pminus 0 & \pminus 0 & \pminus 0 & \pminus 0 & \pminus 0 & \pminus 0 & \pminus 0 & \pminus 0 & \pminus 0 & \pminus 0 & \pminus 0 & \pminus 0 \\
 \pminus 0 & \pminus 0 & \pminus 0 & \pminus 0 & \pminus 0 & \pminus 0 & \pminus 0 & \pminus 0 & \pminus 0 & \pminus 0 & \pminus 0 & \pminus 0 & \pminus 0 & \pminus 0 & \pminus 0 & \pminus 0 & \pminus 0 & \pminus 0 & \pminus 0 & \pminus 0 & \pminus 0 & \pminus 0 & \pminus 0 & \pminus 0 & \pminus 0 & \pminus 0 & \pminus 0 & \pminus 0 & \pminus 0 & \pminus 0 & \pminus 0 & \pminus 0 & \pminus 0 & \pminus 0 & \pminus 0 & \pminus 0 \\
 \pminus 0 & \pminus 0 & \pminus 0 & \pminus 0 & \pminus 0 & \pminus 0 & \pminus 0 & \pminus 0 & \pminus 0 & \pminus 0 & \pminus 0 & \pminus 0 & \pminus 0 & \pminus 0 & \pminus 0 & \pminus 0 & \pminus 0 & \pminus 0 & \pminus 0 & \pminus 0 & \pminus 0 & \pminus 0 & \pminus 0 & \pminus 0 & \pminus 0 & \pminus 0 & \pminus 0 & \pminus 0 & \pminus 0 & \pminus 0 & \pminus 0 & \pminus 0 & \pminus 0 & \pminus 0 & \pminus 0 & \pminus 0 \\
\end{array}
\right)
  }\,,
\end{align}

\begin{align}
\MM_{16} = \scalemath{0.55}{
\left(
\begin{array}{cccccccccccccccccccccccccccccccccccc}
 \pminus 0 & \pminus 0 & \pminus 0 & \pminus 0 & \pminus 0 & \pminus 0 & \pminus 0 & \pminus 0 & \pminus 0 & \pminus 0 & \pminus 0 & \pminus 0 & \pminus 0 & \pminus 0 & \pminus 0 & \pminus 0 & \pminus 0 & \pminus 0 & \pminus 0 & \pminus 0 & \pminus 0 & \pminus 0 & \pminus 0 & \pminus 0 & \pminus 0 & \pminus 0 & \pminus 0 & \pminus 0 & \pminus 0 & \pminus 0 & \pminus 0 & \pminus 0 & \pminus 0 & \pminus 0 & \pminus 0 & \pminus 0 \\
 \pminus 0 & \pminus 0 & \pminus 0 & \pminus 0 & \pminus 0 & \pminus 0 & \pminus 0 & \pminus 0 & \pminus 0 & \pminus 0 & \pminus 0 & \pminus 0 & \pminus 0 & \pminus 0 & \pminus 0 & \pminus 0 & \pminus 0 & \pminus 0 & \pminus 0 & \pminus 0 & \pminus 0 & \pminus 0 & \pminus 0 & \pminus 0 & \pminus 0 & \pminus 0 & \pminus 0 & \pminus 0 & \pminus 0 & \pminus 0 & \pminus 0 & \pminus 0 & \pminus 0 & \pminus 0 & \pminus 0 & \pminus 0 \\
 \pminus 0 & \pminus 0 & \pminus 0 & \pminus 0 & \pminus 0 & \pminus 0 & \pminus 0 & \pminus 0 & \pminus 0 & \pminus 0 & \pminus 0 & \pminus 0 & \pminus 0 & \pminus 0 & \pminus 0 & \pminus 0 & \pminus 0 & \pminus 0 & \pminus 0 & \pminus 0 & \pminus 0 & \pminus 0 & \pminus 0 & \pminus 0 & \pminus 0 & \pminus 0 & \pminus 0 & \pminus 0 & \pminus 0 & \pminus 0 & \pminus 0 & \pminus 0 & \pminus 0 & \pminus 0 & \pminus 0 & \pminus 0 \\
 \pminus 0 & \pminus 0 & \pminus 0 & \pminus 0 & \pminus 0 & \pminus 0 & \pminus 0 & \pminus 0 & \pminus 0 & \pminus 0 & \pminus 0 & \pminus 0 & \pminus 0 & \pminus 0 & \pminus 0 & \pminus 0 & \pminus 0 & \pminus 0 & \pminus 0 & \pminus 0 & \pminus 0 & \pminus 0 & \pminus 0 & \pminus 0 & \pminus 0 & \pminus 0 & \pminus 0 & \pminus 0 & \pminus 0 & \pminus 0 & \pminus 0 & \pminus 0 & \pminus 0 & \pminus 0 & \pminus 0 & \pminus 0 \\
\end{array}
\right)
  }\,,
\end{align}

\begin{align}
\MM_{17} = \scalemath{0.55}{
\left(
\begin{array}{cccccccccccccccccccccccccccccccccccc}
 \pminus 0 & \pminus 0 & \pminus 0 & \pminus 0 & \pminus 0 & \pminus 0 & \pminus 0 & \pminus 0 & \pminus 0 & \pminus 0 & \pminus 0 & \pminus 0 & \pminus 0 & \pminus 0 & \pminus 0 & \pminus 0 & \pminus 0 & \pminus 0 & \pminus 0 & \pminus 0 & \pminus 0 & \pminus 0 & \pminus 0 & \pminus 0 & \pminus 0 & \pminus 0 & \pminus 0 & \pminus 0 & \pminus 0 & \pminus 0 & \pminus 0 & \pminus 0 & \pminus 0 & \pminus 0 & \pminus 0 & \pminus 0 \\
 \pminus 0 & \pminus 0 & \pminus 0 & \pminus 0 & \pminus 0 & \pminus 0 & \pminus 0 & \pminus 0 & \pminus 0 & \pminus 0 & \pminus 0 & \pminus 0 & \pminus 0 & \pminus 0 & \pminus 0 & \pminus 0 & \pminus 0 & \pminus 0 & \pminus 0 & \pminus 0 & \pminus 0 & \pminus 0 & \pminus 0 & \pminus 0 & \pminus 0 & \pminus 0 & \pminus 0 & \pminus 0 & \pminus 0 & \pminus 0 & \pminus 0 & \pminus 0 & \pminus 0 & \pminus 0 & \pminus 0 & \pminus 0 \\
 \pminus 0 & \pminus 0 & \pminus 0 & \pminus 0 & \pminus 0 & \pminus 0 & \pminus 0 & \pminus 0 & \pminus 0 & \pminus 0 & \pminus 0 & \pminus 0 & \pminus 0 & \pminus 0 & \pminus 0 & \pminus 0 & \pminus 0 & \pminus 0 & \pminus 0 & \pminus 0 & \pminus 0 & \pminus 0 & \pminus 0 & \pminus 0 & \pminus 0 & \pminus 0 & \pminus 0 & \pminus 0 & \pminus 0 & \pminus 0 & \pminus 0 & \pminus 0 & \pminus 0 & \pminus 0 & \pminus 0 & \pminus 0 \\
 \pminus 0 & \pminus 0 & \pminus 0 & \pminus 0 & \pminus 0 & \pminus 0 & \pminus 0 & \pminus 0 & \pminus 0 & \pminus 0 & \pminus 0 & \pminus 0 & \pminus 0 & \pminus 0 & \pminus 0 & \pminus 0 & \pminus 0 & \pminus 0 & \pminus 0 & \pminus 0 & \pminus 0 & \pminus 0 & \pminus 0 & \pminus 0 & \pminus 0 & \pminus 0 & \pminus 0 & \pminus 0 & \pminus 0 & \pminus 0 & \pminus 0 & \pminus 0 & \pminus 0 & \pminus 0 & \pminus 0 & \pminus 0 \\
\end{array}
\right)
  }\,,
\end{align}

and $(\MM_{12})_{32,32}=-1$ is the only non vanishing entry in $\MM_{12}$.
 	

\bibliographystyle{JHEP}
\bibliography{references}

\end{document}